\input vanilla.sty
\scaletype{\magstep1}
\hsize = 15truecm

\title FREE QUANTUM FIELDS ON THE POINCAR\'E GROUP. \endtitle

\author                  M. Toller                       \\
\it   Dipartimento di Fisica dell'Universit\`a, Trento,  \\
\it       Istituto Nazionale di Fisica Nucleare,
            Gruppo collegato di Trento, Italia.     \endauthor

\proclaim {Summary} \sl
A class of free quantum fields defined on the Poincar\'e group, is described
by means of their two-point vacuum expectation values.
They are not equivalent to fields defined on the Minkowski spacetime and
they are ``elementary'' in the sense that they
describe particles that transform according to irreducible
unitary representations of the symmetry group, given by the product
of the Poincar\'e group and of the group $SL(2, \bold C)$ considered as an
internal symmetry group.
Some of these fields describe particles with positive mass and
arbitrary spin and particles with zero mass and arbitrary helicity or with an
infinite helicity spectrum. In each case
the allowed $SL(2, \bold C)$ internal quantum numbers are specified.
The properties of local commutativity and the limit in which one recovers
the usual field theories in Minkowski spacetime are discussed.
By means of a superposition of elementary fields, one obtains an example
of a field that present a broken symmetry with respect to
the group $Sp(4, \bold R)$, that survives in the short-distance limit.
Finally, the interaction with an accelerated external source is studied and
and it is shown that, in some theories, the average number of 
particles emitted
per unit of proper time diverges when the acceleration exceeds a finite
critical value.

\rm \endproclaim

\smallpagebreak

\flushpar PACS. 11.10.Kk  -- Field theories in higher dimensions.
\flushpar PACS. 02.20.+b  -- Group theory.

\vfill \eject

\subheading {I. Introduction}

Quantum field theories defined on the Poincar\'e group manifold $\Cal P$
instead of the Minkowski spacetime have been introduced by Lur\c cat$^1$
in 1964.  A motivation of
these investigations was a symmetric treatment of translations, rotations and
Lorentz boosts, namely of all the restricted Poincar\'e transformations.
Later$^{2, 3}$ it has been recognized that
this point of view, in order to be really consistent,
requires a symmetric treatment of velocity,
angular velocity and acceleration; since in relativistic theories there is
an upper bound to the velocity of material objects, one has to introduce
similar limitations to angular velocity and acceleration.  The existence of an
upper bound to the proper acceleration has also been suggested in ref.\ 4.
Brandt$^{5, 6}$ has shown that a maximal acceleration of the order of
$c^2 l_P^{-1}$, where $l_P$ is the Planck length, is expected as
a quantum gravitational effect.

The theories studied in ref.\ 1 are symmetric with respect to both the
left and the right translations of the group $\Cal P$. We suggest that
the physical symmetry group is smaller, namely it contains all the left
translations, but only the
right translations generated by the homogeneous Lorentz group (or its
universal covering $SL(2, \bold C)$). The aim of the present paper
is to analyze the free fields on $\Cal P$ (or on its universal covering
$\tilde{\Cal P}$), that satisfy this weaker symmetry requirement, besides
the natural positivity and spectral conditions.
We begin by considering ``elementary'' fields, namely fields
that describe particles that transform according to irreducible unitary
representations (i. u. r.'s) of the symmetry group. Other fields can be
obtained by superposition of elementary fields, which could provide the
building blocks for the construction of theories implementing
the physical ideas indicated above.

 We do not claim to have found all the fields with the assumed
properties. A formal proof
of this statement would require a more precise formulation of the problem,
for instance a specification of the distribution space to which the fields
have to belong. We think that in a first approach it is more useful to
find as many examples as possible and to introduce technical assumptions
only when they are found to be necessary.

In order to clarify the physical meaning of our assumptions,
it is convenient to consider the theories on an arbitrary group manifold
as special cases of theories of a much larger class$^{7-10}$,
 based on a $n$-dimensional differentiable
manifold $\Cal S$ endowed with a geometric structure defined by $n$ vector
fields $A_{\alpha} \,\,\, (\alpha = 0, \ldots , n-1)$ linearly independent
at every point of $\Cal S$.
If $\Cal S = \Cal G$ is a group manifold, the vector fields $A_{\alpha}$
are the generators of the right translations (invariant under left
translations)$^{11}$
and form a basis of the Lie algebra $\bold l$ of $\Cal G$.

In another interesting case, $\Cal S$
is the 10-dimensional principal bundle$^{12}$ of the pseudo-orthonormal
frames (tetrads) of a (3+1)-dimensional pseudo-Riemannian spacetime $\Cal M$.
The fields
$$ A_4 =A_{23}, \quad A_5=A_{31}, \quad A_6=A_{12},
\quad A_7=A_{10}, \quad A_8=A_{20}, \quad A_9=A_{30} \tag 1.1 $$
are the generators of the structural group of the bundle, 
namely of the Lorentz
group acting on the tetrads. They define a basis of the ``vertical'' subspaces
of the tangent spaces of $\Cal S$. The fields $A_0, \ldots, A_3$ describe the
infinitesimal parallel displacements along the tetrad vectors, namely they
generate the ``horizontal'' subspaces which define a connection.
If $\Cal M$ is the flat Minkowski spacetime, one can identify the bundle of
frames $\Cal S$ with the Poincar\'e group. In a similar way the de Sitter
(or anti-de Sitter) group can be identified with the bundle of frames $\Cal S$
of a pseudo-Riemannian spacetime $\Cal M$ with constant positive
(or negative) curvature.

Classical field theories including gravitation
based on these ideas have been developed in refs.\ 8-10, 13, 14.
If one identifies $\Cal S$ with a principal bundle with a larger
structural group, one can also treat Maxwell and Yang-Mills fields$^{8, 15}$.
Any classical or quantum field
defined on the pseudo-Riemannian space-time $\Cal M$ can easily be
translated into a field defined on the bundle of frames $\Cal S$.
We think, however, that the new formalism should be used to formulate
new physical ideas.

The fields $A_{\alpha}$ define a ``teleparallelism'' in $\Cal S$,
namely a set of isomorphisms between all
the tangent spaces of $\Cal S$ and a fixed vector space $\Cal T = \bold R^n$.
A closed wedge $\Cal T^+ \subset \Cal T$ defines a field of
wedges in the tangent spaces of $\Cal S$ which describes the causal properties
of the theory and in particular the upper bounds to velocity, angular velocity
and acceleration$^{2, 3}$.

It is useful  to introduce the  structure coefficients
$F_{\alpha \beta}^{\gamma}$ defined by
$$ [A_{\alpha}, A_{\beta}] = F_{\alpha \beta}^{\gamma} A_{\gamma}, \tag 1.2 $$
where $[A, B]$ is the Lie bracket of two vector fields. If $\Cal S$ is a group
manifold, they are the structure constants of the corresponding Lie algebra.
If $\Cal S$ is a bundle of frames, some of the structure coefficients  are
the components of the curvature and torsion tensors and in a theory of
gravitation they have a dynamical character, namely they depend on the
distribution of matter.

In accord with the ideas indicated above,
it is natural to consider theories in which all the structure coefficients
$F_{\alpha \beta}^{\gamma}$ and all the vector fields
$A_{\alpha}$ have a dynamical character.
In a theory of this kind, the field equations
determine both the fields $A_{\alpha}$
that describe the geometry and the  fields $\psi_{\rho}$ that describe matter.
We assume that this theory is invariant under all the diffeomorphisms of
$\Cal S$ and under a symmetry group $\Cal F$ acting linearly in the
following way
$$ \psi_{\rho} \to S_{\rho}{}^{\sigma}(k) \psi_{\sigma}, \tag 1.3 $$
$$ A_{\alpha} \to C_{\alpha}{}^{\beta}(k) A_{\beta},  \qquad k \in \Cal F,
\tag 1.4 $$
where $S$ and $C$ are linear representations of $\Cal F$. The representation
$C$ is real and one can consider it as acting on the vector space $\Cal T$.
It is natural to require that the transformations $C(k)$ leave
the wedge $\Cal T^+$ invariant.
It follows from the physical interpretation of the fields $A_{\alpha}$
that the element $k$ cannot depend on the point of $\Cal S$, namely
$\Cal F$ is a global symmetry group.

In the physically most interesting example, $\Cal S$ is a ten-dimensional
manifold, the group $\Cal F$ contains a subgroup
isomorphic to  the restricted Lorentz group
$SO^{\uparrow}(3, 1)$ and the representation $C$ restricted to this
subgroup is the direct sum of the vector and the antisymmetric tensor
representations. Actually, in order to treat matter fields with
half-integral spin, it is convenient to assume that $\Cal F$ contains
the universal covering of $SO^{\uparrow}(3, 1)$, namely $SL(2, \bold C)$.
Under these conditions, if we assume that
$\Cal T^+$ is a cone (namely that $\Cal T^+ \cap - \Cal T^+ = \{0\}$)
with interior points, this cone is determined up to a change of the
units of time and length$^{2, 3}$. It has a large symmetry group
given by $L(4, \bold R)$ acting on $\Cal T$ by means of its symmetric
tensor representation$^2$. In this framework, it is natural to assume that
the symmetry group $\Cal F$ of the theory is  $L(4, \bold R)$  or one
of its subgroups that contain $SL(2, \bold C)$. Possible choices are
$SL(4, \bold R)$, $Sp(4, \bold R)$ or $SL(2, \bold C)$.

In the present paper we consider the fields $A_{\alpha}$ as fixed
classical fields and we concentrate our attention on the quantum
fields that describe matter. The symmetry group of this partial theory
contains only the elements of the symmetry group of the complete theory
that do not affect the geometric fields $A_{\alpha}$.  For instance,
the elements of $\Cal F$ that do not preserve the values of the
structure coefficients $F_{\alpha \beta}^{\gamma}$
represent broken symmetries. If we consider the
values of the structure coefficients as expectation  values
of some fields in a vacuum
state of the complete theory, this is a spontaneous symmetry breaking.

In Sec. II we discuss some general properties of quantized matter fields
on an arbitrary connected Lie group.
In Sec. III we begin the treatment of free quantum fields on the
universal covering $\tilde{\Cal P}$ of the restricted Poincar\'e group
(namely the inhomogeneous $SL(2, \bold C)$ group). The subgroup of $\Cal F$
that survives the symmetry breaking must preserve both the cone $\Cal T^+$
and the structure constants of the Poincar\'e Lie algebra. It follows that
it must coincide with $SL(2, \bold C)$, as we have anticipated above.
The unbroken symmetry group of the theory is the product of $\tilde{\Cal P}$
and $SL(2, \bold C)$, considered as an internal symmetry group.
The free quantum fields are completely described by
the two-point Wightman distributions 
(or vacuum expectation values, v. e. v.'s)
and we give a general representation
of the distributions that satisfy the appropriate symmetry, spectral and
positivity conditions. In Sec. IV we treat the commutation or
anticommutation properties of the free fields and we discuss the connection
between spin and statistics, which is not the usual one. For instance,
a ``scalar'' field, namely a field with only one component, has to be
quantized with commutators, but it can describe particles with
any spin. In Sec. V we treat the positive-mass case with
arbitrary spin and we write explicitly a wide class of v. e. v.'s
in terms of the matrix elements of the i. u. r.'s of $SL(2, \bold C)$.
In Sec. VI we consider the more delicate zero-mass case and we find
theories that describe particles with an infinite helicity spectrum
(not observed in nature) and particles with an arbitrary given helicity.
Scalar fields that describe particles with zero mass and
given nonvanishing helicity have pathological features.

Since the non trivial i. u. r.'s of $SL(2, \bold C)$ are 
infinite-dimensional,
the mass spectrum of these theories is infinitely degenerate. In order
to avoid evident contradictions with the known physical phenomena, we have
to require that, when the mass is within the range of presently available
energies, only a finite number of internal states of the particles can
be excited by the field with appreciable probability. This happens
when the parameters which  label the i. u. r.'s of $SL(2, \bold C)$
approach the limit $ M = j = 0, \,\, c \to 1$. Actually, in this limit
the v. e. v.'s tend to the ones that define the usual scalar
free field in Minkowski spacetime.  This problem is less relevant
when the mass is of the order of the Planck mass. The theories with
a broken higher symmetry satisfy these requirements automatically.

In Sec. VII we find the differential equations satisfied by the quantum
fields defined in the preceding Sections and we compare them with some
field equations in a flat ten-dimensional space. In Sec. VII we introduce
the concept of ``spin-mass-shell'' and we discuss the relation between
the v. e. v.'s on $\tilde{\Cal P}$ and the corresponding
distributions defined on $\bold R^{10}$. We show that not all the free
fields on the flat space have a corresponding field on $\tilde{\Cal P}$.

In the remaining Sections we give some examples, in order to illustrate
the general formalism. A more complete treatment will be given elsewhere.
In Sec. IX we consider
a theory on the flat space symmetric with respect to the group
$Sp(4, \bold R)$  and we build the corresponding theory
on $\tilde{\Cal P}$. In this theory the higher symmetry is broken, but
the v. e. v.'s maintain the higher symmetry in the short-distance
limit. This is an explicit example of a new kind of broken symmetry in
quantum field theory.

In Sec. X we consider an external source, represented by an accelerated
disk, interacting with one of the fields defined in the present paper.
We consider with more detail the field introduced in Sec. IX
and we show that the number of particles emitted per unit of
proper time diverges when the acceleration exceeds a finite critical value.
This result shows that the formalism really contains, in some sense, the
ideas that provided its motivation. Brandt$^{6}$ has suggested that, when
the acceleration of a particle approaches a critical value, 
the energy radiated in the form of quantum black holes diverges, 
preventing a larger acceleration.
The formalism presented here could provide a simplified model for this
process, if the particles described by our fields are interpreted as
quantum black holes$^{16}$. This interpretation, however, raises several
difficult problems.

\bigpagebreak

\subheading {II. Quantum Fields on a Group Manifold}

As we have anticipated in the Introduction, we consider a $n$-dimensional
manifold $\Cal S$ with $n$ vector fields  $A_{\alpha}
\,\, (\alpha = 0,\ldots, n-1)$ that, being linearly independent,
 define a basis in each tangent space of
$\Cal S$. As a consequence, we can identify all the tangent spaces with a
single vector space $\Cal T$.  These vector fields describe
the gravitational field and possibly other gauge fields, while
matter is described by a set of fields $\psi_{\rho}$, on the manifold $\Cal S$
$^{7-10}$. We assume that the complete field equations, including gravitation,
are invariant under all the diffeomorphisms of $\Cal S$ and
under a group $\Cal F$ which acts on the matter fields
and on the geometric fields according to the linear formulas (1.3), (1.4).

We consider the fields
$A_{\alpha}$ as classical external fields and
we  restrict our considerations to symmetry transformations that
leave them invariant. The transformations which have this property
are implemented by unitary or anti-unitary operators
acting on the Hilbert space $\Cal H$ which describes the states of matter.
Since we consider only connected symmetry groups, we deal only with
unitary symmetry operators$^{17}$. Moreover, we
assume that they form a continuous representation of the 
symmetry group$^{18}$.
The fields $\psi_{\rho}$ are operator-valued distributions on $\Cal S$ that
act in a dense linear subspace $\Cal D$ of $\Cal H$. We assume that both the
smeared field operators and the symmetry operators transform $\Cal D$ into
itself.

Field theories on a $n$-dimensional manifold $\Cal S$ which has a symmetry
group have been treated in ref.\ 7. Here we consider the case in
which the vector fields $A_{\alpha}$ generate a connected $n$-dimensional
Lie group $\Cal G$ of diffeomorphisms of $\Cal S$.
It is convenient to assume that
$\Cal G$ acts on the right in the space $\Cal S$, namely the action of
the element $g \in \Cal G$ on the element $s \in \Cal S$ is written
as $(g, s) \to sg$. We assume also that $\Cal G$ acts freely and transitively;
it follows that if we choose an origin $s_0 \in \Cal S$, the mapping
$g \to s = s_0 g$ is a diffeomorphism of $\Cal G$ onto $\Cal S$.
The action of $\Cal G$ on $\Cal S$ takes the form
$$ s = s_0 g \to s' = s_0 g h =  s h, \qquad  g, h \in \Cal G, \tag 2.1 $$
namely it corresponds to a right translation of $\Cal G$.
The vector fields $A_{\alpha}$ can also be considered as
fields on $\Cal G$, that generate the right translations.
They form a basis of the Lie algebra $\bold l$ of $\Cal G$, that,
as a vector space, can be identified with $\Cal T$.

The vector fields $A_{\alpha}$ are invariant under the left translations
of $\Cal G$; it follows that a transformation of the kind
$$ s = s_0 g \to s' = s_0 h g, \qquad g, h \in \Cal G \tag 2.2 $$
is a symmetry transformation that leaves the geometry of $\Cal S$ invariant.
It can be interpreted as a change of the origin $s_0$.

The diffeomorphisms of the kind (2.2) provide a first class of
symmetry transformations. They act on the matter fields in the following way:
$$ \psi_{\rho}(s_0 h^{-1} g) = U(h)^{-1} \psi_{\rho}(s_0 g) U(h),
\qquad g, h \in \Cal G. \tag 2.3 $$
Note that the indices of the fields are not involved: every component behaves
as a scalar field. We indicate by $U(A_{\alpha})$ the generators of the
continuous unitary representation $U$ corresponding to the elements
$A_{\alpha}$ of $\bold l$.  The operators $i U(A_{\alpha})$
are self-adjoint and are interpreted as the energy, the momentum,
the relativistic angular momentum and possibly (if $n>10$) the charges that
generate some gauge transformations. Note that, as it is
expected, these operators, as well as the unitary operators $U(h)$,
depend on the choice of the frame $s_0$.

The diffeomorphisms of the kind (2.1) affect the geometric fields according
to the formula
$$ A_{\alpha} \to A_{\beta} B^{\beta}{}_{\alpha}(h) , \tag 2.4 $$
where $B^{\beta}{}_{\alpha}(h)$ is the adjoint representation of $\Cal G$.
They give rise to symmetry transformations only if eq\. (2.4) can
be compensated by a transformation of the kind (1.4), namely if
$$ B^{\beta}{}_{\alpha}(h) = C_{\alpha}{}^{\beta}(\hat h^{-1}), \qquad
h \in \Cal G, \qquad \hat h \in \Cal F. \tag 2.5 $$
This condition defines a subgroup $\Cal G_2 \subset \Cal G$ and we assume that
$h \to \hat h$ is a continuous homomorphism of $\Cal G_2$ onto a subroup
$\Cal F_2 \subset \Cal F$.
In the following we write $S(h)$ instead of $S(\hat h)$. Then we have
$$ S_{\rho}{}^{\sigma}(h) \psi_{\sigma}(s h) = V(h)^{-1} \psi_{\rho}(s) V(h),
\qquad  h \in \Cal G_2. \tag 2.6 $$
The operators $V(h)$ commute with the operators $U(g)$ and do not depend
on the choice of $s_0$. They describe a kind of internal symmetry.

The internal automorphisms of $\Cal G$, given by $g \to hgh^{-1}$,
are the product of a right and a left translation and do not require
a separate treatment. The external automorphisms of $\Cal G$ (for
instance the spacetime dilatations in the Poincar\'e group) give
rise to a new kind of symmetry if their action on the fields
$A_{\alpha}$ can be compensated by a transformation of the kind (1.4).
This compensation is not possible in the cases we shall treat in the
following Sections.

Several general features of the quantum field theories on Minkowski spacetime
$^{19, 20}$ can be extended to the theories on a group manifold.
We assume that there is a vacuum state $\Omega \in \Cal D$ invariant with
respect to both the representations $U$ and $V$, and we define the v. e. v.
$$ (\Omega, \psi_{\rho}(s_1) \psi_{\sigma}(s_2) \Omega) =
\Cal W_{\rho \sigma}(s_1, s_2). \tag 2.7 $$
It follows from eq\. (2.3) that this quantity can be considered as a
distribution on $\Cal G$. In fact, we have
$$ \Cal W_{\rho \sigma}(s_1, s_2) = \Cal W_{\rho \sigma}(s_0 g_1, s_0 g_2) =
W_{\rho \sigma}(g_1^{-1} g_2). \tag 2.8 $$
In the following we understand a fixed choice of $s_0$ and we write
$\psi_{\rho}(g)$ instead of $\psi_{\rho}(s_0 g)$.
The v. e. v. of $m+1$ fields is defined in the following
way as a distribution on $\Cal G^m$:
$$ (\Omega, \psi_{\rho_1}(g_1) \cdots \psi_{\rho_{m+1}}(g_{m+1}) \Omega) =
 W^{(m)}_{\rho_1 \ldots \rho_{m+1}}(g_1^{-1} g_2, \ldots,
g_m^{-1} g_{m+1}). \tag 2.9 $$

If the fields $\psi_{\rho}(g)$ are not Hermitean, we have
to consider the Hermitean conjugate as a different field and
we use the notation
$$ \psi_{\rho}^{\dagger}(g) = \psi_{\overline\rho}(g), \qquad
S_{\overline\rho}{}^{\overline\sigma}(a) = \overline S_{\rho}{}^{\sigma}(a).
\tag 2.10 $$
In all the formulas any index can be replaced by a barred index and
vice versa, unless it is stated otherwise. If the field is Hermitean,
we have to put $\overline\rho = \rho$ and the representation $S$ must be real.
Then from the definition we get
$$ \overline W^{(m)}_{\rho_1 \ldots \rho_{m+1}}(g_1, \ldots, g_m) =
  W^{(m)}_{\overline\rho_{m+1} \ldots \overline\rho_1}
(g_m^{-1}, \ldots, g_1^{-1}) \tag 2.11 $$
and from eq\. (2.6) we get the symmetry property
$$ S_{\rho_1}{}^{\sigma_1}(h) \cdots S_{\rho_{m+1}}{}^{\sigma_{m+1}}(h) \,
W^{(m)}_{\sigma_1 \ldots \sigma_{m+1}}(h^{-1} g_1 h, \ldots, h^{-1} g_m h) =
\tag 2.12  \nopagebreak $$
$$ = W^{(m)}_{\rho_1 \ldots \rho_{m+1}}(g_1, \ldots, g_m),
\qquad h \in \Cal G_2. $$

If we deal with free fields, all the v. e. v.'s can be
obtained from the two-point distributions by means of the Wick theorem.
In this case, the vectors of the kind
$$ \Phi(f) = \int f^{\rho}(g) \psi^{\dagger}_{\rho}(g) \, dg \,\, \Omega,
\tag 2.13 $$
where $f$ is a test function and $dg$ is an invariant measure on $\Cal G$,
are dense in the Hilbert space $\Cal H^{(1)}$ of the ``one-particle'' states.
The square of the norm of a the vector (2.13) is given by
$$ (\Phi(f), \Phi(f)) = \int \overline{f^{\rho}}(g_1) f^{\sigma}(g_2)
  W_{\rho \overline\sigma}(g_1^{-1} g_2)\, dg_1\, dg_2  \geq 0.
\tag 2.14 $$
This is the positivity condition.
For interacting field theories, we have more complicated positivity
conditions that involve all the other v. e. v.'s.

The symmetry operators defined in eqs\. (2.3) and (2.6) act on the
one-particle states in the following way:
$$ U^{(1)}(h) \Phi(f) = \Phi(f'),
\qquad f^{\prime \rho}(g) =  f^{\rho}(h^{-1} g), \tag 2.15 $$
$$ V^{(1)}(h) \Phi(f) = \Phi(f'),
\qquad f^{\prime \rho}(g) = f^{\sigma}(g h)
\overline S_{\sigma}{}^{\rho}(h^{-1}). \tag 2.16 $$
Eqs\. (2.14)-(2.16) permit, in the usual way,
the reconstruction of the one-particle Hilbert space (as the completion
of a quotient) and of the operators $U^{(1)}(h)$ and $V^{(1)}(h)$.
If the unitary representation $U^{(1)} \times V^{(1)}$ of
$\Cal G \times \Cal G_2$ is irreducible, we say that the free field
is ``elementary''.

\subheading {II. Quantum Fields on a Group Manifold}

As we have anticipated in the Introduction, we consider a $n$-dimensional
manifold $\Cal S$ with $n$ vector fields  $A_{\alpha}
\,\, (\alpha = 0,\ldots, n-1)$ that, being linearly independent,
 define a basis in each tangent space of
$\Cal S$. As a consequence, we can identify all the tangent spaces with a
single vector space $\Cal T$.  These vector fields describe
the gravitational field and possibly other gauge fields, while
matter is described by a set of fields $\psi_{\rho}$, on the manifold $\Cal S$
$^{7-10}$. We assume that the complete field equations, including gravitation,
are invariant under all the diffeomorphisms of $\Cal S$ and
under a group $\Cal F$ which acts on the matter fields
and on the geometric fields according to the linear formulas (1.3), (1.4).

We consider the fields
$A_{\alpha}$ as classical external fields and
we  restrict our considerations to symmetry transformations that
leave them invariant. The transformations which have this property
are implemented by unitary or anti-unitary operators
acting on the Hilbert space $\Cal H$ which describes the states of matter.
Since we consider only connected symmetry groups, we deal only with
unitary symmetry operators$^{17}$. Moreover, we
assume that they form a continuous representation of the 
symmetry group$^{18}$.
The fields $\psi_{\rho}$ are operator-valued distributions on $\Cal S$ that
act in a dense linear subspace $\Cal D$ of $\Cal H$. 
We assume that both the
smeared field operators and the symmetry operators transform $\Cal D$ into
itself.

Field theories on a $n$-dimensional manifold $\Cal S$ which has a symmetry
group have been treated in ref.\ 7. Here we consider the case in
which the vector fields $A_{\alpha}$ generate a connected $n$-dimensional
Lie group $\Cal G$ of diffeomorphisms of $\Cal S$.
It is convenient to assume that
$\Cal G$ acts on the right in the space $\Cal S$, namely the action of
the element $g \in \Cal G$ on the element $s \in \Cal S$ is written
as $(g, s) \to sg$. We assume also that $\Cal G$ acts 
freely and transitively;
it follows that if we choose an origin $s_0 \in \Cal S$, the mapping
$g \to s = s_0 g$ is a diffeomorphism of $\Cal G$ onto $\Cal S$.
The action of $\Cal G$ on $\Cal S$ takes the form
$$ s = s_0 g \to s' = s_0 g h =  s h, \qquad  g, h \in \Cal G, \tag 2.1 $$
namely it corresponds to a right translation of $\Cal G$.
The vector fields $A_{\alpha}$ can also be considered as
fields on $\Cal G$, that generate the right translations.
They form a basis of the Lie algebra $\bold l$ of $\Cal G$, that,
as a vector space, can be identified with $\Cal T$.

The vector fields $A_{\alpha}$ are invariant under the left translations
of $\Cal G$; it follows that a transformation of the kind
$$ s = s_0 g \to s' = s_0 h g, \qquad g, h \in \Cal G \tag 2.2 $$
is a symmetry transformation that leaves the geometry of $\Cal S$ invariant.
It can be interpreted as a change of the origin $s_0$.

The diffeomorphisms of the kind (2.2) provide a first class of
symmetry transformations. They act on the matter fields in the following way:
$$ \psi_{\rho}(s_0 h^{-1} g) = U(h)^{-1} \psi_{\rho}(s_0 g) U(h),
\qquad g, h \in \Cal G. \tag 2.3 $$
Note that the indices of the fields are not involved: every component behaves
as a scalar field. We indicate by $U(A_{\alpha})$ the generators of the
continuous unitary representation $U$ corresponding to the elements
$A_{\alpha}$ of $\bold l$.  The operators $i U(A_{\alpha})$
are self-adjoint and are interpreted as the energy, the momentum,
the relativistic angular momentum and possibly (if $n>10$) the charges that
generate some gauge transformations. Note that, as it is
expected, these operators, as well as the unitary operators $U(h)$,
depend on the choice of the frame $s_0$.

The diffeomorphisms of the kind (2.1) affect the geometric fields according
to the formula
$$ A_{\alpha} \to A_{\beta} B^{\beta}{}_{\alpha}(h) , \tag 2.4 $$
where $B^{\beta}{}_{\alpha}(h)$ is the adjoint representation of $\Cal G$.
They give rise to symmetry transformations only if eq\. (2.4) can
be compensated by a transformation of the kind (1.4), namely if
$$ B^{\beta}{}_{\alpha}(h) = C_{\alpha}{}^{\beta}(\hat h^{-1}), \qquad
h \in \Cal G, \qquad \hat h \in \Cal F. \tag 2.5 $$
This condition defines a subgroup $\Cal G_2 \subset \Cal G$ and we assume that
$h \to \hat h$ is a continuous homomorphism of $\Cal G_2$ onto a subroup
$\Cal F_2 \subset \Cal F$.
In the following we write $S(h)$ instead of $S(\hat h)$. Then we have
$$ S_{\rho}{}^{\sigma}(h) \psi_{\sigma}(s h) = V(h)^{-1} \psi_{\rho}(s) V(h),
\qquad  h \in \Cal G_2. \tag 2.6 $$
The operators $V(h)$ commute with the operators $U(g)$ and do not depend
on the choice of $s_0$. They describe a kind of internal symmetry.

The internal automorphisms of $\Cal G$, given by $g \to hgh^{-1}$,
are the product of a right and a left translation and do not require
a separate treatment. The external automorphisms of $\Cal G$ (for
instance the spacetime dilatations in the Poincar\'e group) give
rise to a new kind of symmetry if their action on the fields
$A_{\alpha}$ can be compensated by a transformation of the kind (1.4).
This compensation is not possible in the cases we shall treat in the
following Sections.

Several general features of the quantum field theories on Minkowski spacetime
$^{19, 20}$ can be extended to the theories on a group manifold.
We assume that there is a vacuum state $\Omega \in \Cal D$ invariant with
respect to both the representations $U$ and $V$, and we define the v. e. v.
$$ (\Omega, \psi_{\rho}(s_1) \psi_{\sigma}(s_2) \Omega) =
\Cal W_{\rho \sigma}(s_1, s_2). \tag 2.7 $$
It follows from eq\. (2.3) that this quantity can be considered as a
distribution on $\Cal G$. In fact, we have
$$ \Cal W_{\rho \sigma}(s_1, s_2) = \Cal W_{\rho \sigma}(s_0 g_1, s_0 g_2) =
W_{\rho \sigma}(g_1^{-1} g_2). \tag 2.8 $$
In the following we understand a fixed choice of $s_0$ and we write
$\psi_{\rho}(g)$ instead of $\psi_{\rho}(s_0 g)$.
The v. e. v. of $m+1$ fields is defined in the following
way as a distribution on $\Cal G^m$:
$$ (\Omega, \psi_{\rho_1}(g_1) \cdots \psi_{\rho_{m+1}}(g_{m+1}) \Omega) =
 W^{(m)}_{\rho_1 \ldots \rho_{m+1}}(g_1^{-1} g_2, \ldots,
g_m^{-1} g_{m+1}). \tag 2.9 $$

If the fields $\psi_{\rho}(g)$ are not Hermitean, we have
to consider the Hermitean conjugate as a different field and
we use the notation
$$ \psi_{\rho}^{\dagger}(g) = \psi_{\overline\rho}(g), \qquad
S_{\overline\rho}{}^{\overline\sigma}(a) = \overline S_{\rho}{}^{\sigma}(a).
\tag 2.10 $$
In all the formulas any index can be replaced by a barred index and
vice versa, unless it is stated otherwise. If the field is Hermitean,
we have to put $\overline\rho = \rho$ and the representation $S$ must be real.
Then from the definition we get
$$ \overline W^{(m)}_{\rho_1 \ldots \rho_{m+1}}(g_1, \ldots, g_m) =
  W^{(m)}_{\overline\rho_{m+1} \ldots \overline\rho_1}
(g_m^{-1}, \ldots, g_1^{-1}) \tag 2.11 $$
and from eq\. (2.6) we get the symmetry property
$$ S_{\rho_1}{}^{\sigma_1}(h) \cdots S_{\rho_{m+1}}{}^{\sigma_{m+1}}(h) \,
W^{(m)}_{\sigma_1 \ldots \sigma_{m+1}}(h^{-1} g_1 h, \ldots, h^{-1} g_m h) =
\tag 2.12  \nopagebreak $$
$$ = W^{(m)}_{\rho_1 \ldots \rho_{m+1}}(g_1, \ldots, g_m),
\qquad h \in \Cal G_2. $$

If we deal with free fields, all the v. e. v.'s can be
obtained from the two-point distributions by means of the Wick theorem.
In this case, the vectors of the kind
$$ \Phi(f) = \int f^{\rho}(g) \psi^{\dagger}_{\rho}(g) \, dg \,\, \Omega,
\tag 2.13 $$
where $f$ is a test function and $dg$ is an invariant measure on $\Cal G$,
are dense in the Hilbert space $\Cal H^{(1)}$ of the ``one-particle'' states.
The square of the norm of a the vector (2.13) is given by
$$ (\Phi(f), \Phi(f)) = \int \overline{f^{\rho}}(g_1) f^{\sigma}(g_2)
  W_{\rho \overline\sigma}(g_1^{-1} g_2)\, dg_1\, dg_2  \geq 0.
\tag 2.14 $$
This is the positivity condition.
For interacting field theories, we have more complicated positivity
conditions that involve all the other v. e. v.'s.

The symmetry operators defined in eqs\. (2.3) and (2.6) act on the
one-particle states in the following way:
$$ U^{(1)}(h) \Phi(f) = \Phi(f'),
\qquad f^{\prime \rho}(g) =  f^{\rho}(h^{-1} g), \tag 2.15 $$
$$ V^{(1)}(h) \Phi(f) = \Phi(f'),
\qquad f^{\prime \rho}(g) = f^{\sigma}(g h)
\overline S_{\sigma}{}^{\rho}(h^{-1}). \tag 2.16 $$
Eqs\. (2.14)-(2.16) permit, in the usual way,
the reconstruction of the one-particle Hilbert space (as the completion
of a quotient) and of the operators $U^{(1)}(h)$ and $V^{(1)}(h)$.
If the unitary representation $U^{(1)} \times V^{(1)}$ of
$\Cal G \times \Cal G_2$ is irreducible, we say that the free field
is ``elementary''.

\bigpagebreak
 
\subheading {III. Free Fields on the Poincar\'e Group}

Now we apply the general results of the preceding Section to the Poincar\'e
group. We consider the ten-dimensional manifold $\Cal S$ of all the Lorentz
reference frames in the Minkowski space $\Cal M$ which are left-handed
and future-directed.  If we fix a reference frame $s_0$,
for each reference frame $s$ there is one and only one element
of the proper orthochronous Poincar\'e group $\Cal P$ that transforms  $s_0$
into $s$. Then we can identify the space $\Cal S$ with  $\Cal P$ and
consider quantum fields defined on it. Actually, in order to treat fields
with half-integral spin, it is convenient to use the universal covering
$\tilde{\Cal P}$ of $\Cal P$, namely the semidirect
product of the four-dimensional translation group $\bold R^4$ and
$SL(2, \bold C)$. For the elements of this group and for their
multiplication rule, we use the standard notation
$$ (x, a) (y, b) = (x + \Lambda (a)y, ab), \qquad x, y \in \bold R^4, \quad
a, b \in SL(2, \bold C), \tag 3.1 $$
where $\Lambda (a)$ is the $4\times 4$ Lorentz matrix corresponding to the
element $a \in SL(2, \bold C)$.  For the scalar product of two
four-vectors we use the notation
$$ g_{ik} x^i y^k = x \cdot y = -x^0 y^0 + \bold x \cdot \bold y, \qquad
\bold x \cdot \bold y = x^1 y^1 + x^2 y^2 + x^3 y^3. \tag 3.2 $$
The one-parameter subgroups of $SL(2, \bold C)$ that correspond
to rotations around the axes and to pure Lorentz transformations
along the axes are written as
$$ u_k(\theta) = \exp(-\tfrac 1 2 i \theta \sigma_k),  \qquad
   a_k(\zeta) = \exp(\tfrac 1 2 \zeta \sigma_k),  \tag 3.3 $$
where $\sigma_k$ are the Pauli matrices.

We assume that the group $\Cal F$ contains $SL(2, \bold C)$ and that it
preserves the cone $\Cal T^+$ defined in refs. 2, 3. It is easy to see
that the translations (acting on $\Cal T$ by means of the
adjoint representation of $\Cal P$)
and the spacetime dilatations (which are external
automorphisms of $\tilde{\Cal P}$ and of the corresponding Lie algebra)
do not preserve the cone $\Cal T^+$.  It follows that we have to put
$\Cal G_2 = SL(2, \bold C)$.
Then the general equations (2.3) and (2.6) take the form
$$ \psi_{\rho}((y, b)^{-1}(x,a)) =
\psi_{\rho}(\Lambda(b)^{-1}(x - y), b^{-1}a) =
U(y, b)^{-1} \psi_{\rho}(x, a) U(y, b),  \tag 3.4 $$
$$ S_{\rho}{}^{\sigma}(b) \psi_{\sigma}(x, ab) =
V(b)^{-1} \psi_{\rho}(x, a) V(b). \tag 3.5 $$

If $V(b) = 1$, from eq\. (3.5) we have
$$\psi_{\rho}(x, a) = S_{\rho}{}^{\sigma}(a^{-1}) \psi_{\sigma}(x, 1)
 \tag 3.6 $$
and we are dealing with a field $\psi_{\sigma}(x) = \psi_{\sigma}(x, 1)$
defined on the Minkowski space-time. From eq\. (3.4), we get the usual
covariance property
$$  S_{\rho}{}^{\sigma}(b) \psi_{\sigma}(\Lambda(b)^{-1}(x - y)) =
U(y, b)^{-1} \psi_{\rho}(x) U(y, b).  \tag 3.7 $$
In the following we consider the case in which $V(b)$ is not trivial.

A free field theory is completely described by the two-point v. e. v.'s
$$ (\Omega, \psi_{\rho}(x, a) \psi_{\sigma}(y, b) \Omega) =
W_{\rho \sigma}((x, a)^{-1} (y, b )) =
W_{\rho \sigma}(\Lambda(a^{-1})(y-x), a^{-1} b). \tag 3.8 $$
From the covariance property (3.5) we get the formula
$$ W_{\rho \sigma}(\Lambda(b)x, b a b^{-1}) =
S_{\rho}{}^{\mu}(b) S_{\sigma}{}^{\nu}(b) W_{\mu \nu}(x, a)
\tag 3.9 $$
(remember our conventions about the introduction of barred indices).
If the field is not Hermitean, we assume
$$ W_{\rho \sigma}(x, a) = W_{\overline\rho \overline\sigma}(x, a) = 0
\tag 3.10 $$
(here the bars over the indices cannot be modified)
and the distributions $ W_{\rho \overline\sigma}(x, a)$ and
$W_{\overline\rho \sigma}(x, a)$ can be treated independently unless
we introduce some requirement of local commutativity (see the next Section).

The states of the form
$$\Phi(f) = \int f^{\rho}(x, a) \psi^{\dagger}_{\rho}(x, a) \,
d^4 x \, d^6 a \,\, \Omega, \tag 3.11 $$
where $f$ is an arbitrary test function with compact support and $d^6 a$
is an invariant measure on $SL(2, \bold C)$,
form a dense set in the Hilbert subspace $\Cal H ^{(1)}$ of the one-particle
states.  Their norm is given by the formula
$$ (\Phi(f), \Phi(f)) = \tag 3.12 \nopagebreak $$
$$ = \int \overline{f^{\rho}}(x, a) f^{\sigma}(y, b)
W_{\rho \overline\sigma}(\Lambda(a^{-1})(y-x), a^{-1} b)
\, d^4 x \, d^6 a \, d^4 y \, d^6 b  \geq 0, $$
that also gives the positivity property of the two-point distribution.

We want to find the solutions of the conditions (3.9) and (3.12) that
describe elementary free fields. More general solutions can be obtained
by means of sums or integrals of these solutions.
If we introduce the Fourier transformations
$$ W_{\rho \overline\sigma}(x, a) = \int \exp(-i k \cdot x)
\tilde W_{\rho \overline\sigma}(k, a) \, d^4 k, \tag 3.13 $$
$$ \tilde f^{\rho}(k, a) = \int \exp(-i k \cdot x) f^{\rho}(x, a)
 \, d^4 x, \tag 3.14 $$
eq\. (3.12) takes the form
$$ (\Phi(f), \Phi(f)) = \int \overline{{\tilde f}^{\rho}}(k, a)
\tilde f^{\sigma}(k, b)
\tilde W_{\rho \overline\sigma}(\Lambda(a^{-1}) k, a^{-1} b)
\, d^6 a \, d^6 b \, d^4 k  \geq 0  \tag 3.15 $$
and eq. (3.9) gives
$$ \tilde W_{\rho \overline\sigma}(\Lambda(b) k, b a b^{-1}) =
 S_{\rho}{}^{\mu}(b) \overline S_{\sigma}{}^{\nu}(b)
\tilde W_{\mu \overline\nu}(k, a). \tag 3.16 $$

We assume that the distribution $\tilde W_{\rho \overline\sigma}(k, a)$
vanishes if $k$ does not
belong to the future cone (spectral condition). Since the elementary fields
have a definite mass $\mu$, this distribution
has support on the orbit defined by
$$  k \cdot k = - \mu^2, \qquad k^0 > 0  \tag 3.17 $$
and has the form
$$  \tilde W_{\rho \overline\sigma}(k, a) = (2 \pi)^{-3}
w_{\rho \overline\sigma}(k, a)
\theta(k^0) \delta(k \cdot k + \mu^2), \tag 3.18 $$
where $\theta$ is the step function.

Following a procedure introduced by Wigner$^{21}$, we choose a representative
element on each orbit
$$ \hat k = (\mu, 0, 0, 0), \qquad \mu > 0, \tag 3.19 $$
$$ \hat k = (1, 0, 0, 1), \qquad \mu = 0 \tag 3.20 $$
and for each value of the four-momentum $k$ on the orbit, we choose an element
$a_k \in SL(2, \bold C)$ with the property
$$ k = \Lambda(a_k) \hat k. \tag 3.21 $$
Then we see from eq\. (3.16) that we can put
$$ w_{\rho \overline\sigma}(k, a) =
S_{\rho}{}^{\mu}(a_k) \overline S_{\sigma}{}^{\nu}(a^{-1} a_k)
w_{\mu \overline\nu}(a_k^{-1} a a_k),    \tag 3.22 $$
where
$$ w_{\rho \overline\sigma}(a) = \overline S_{\sigma}{}^{\nu}(a)
w_{\rho \overline\nu}(\hat k, a). \tag 3.23 $$
In this way we obtain the integral representation
$$ W_{\rho \overline\sigma}(x, a) = \tag 3.24 \nopagebreak $$
$$ = (2 \pi)^{-3} \int \exp(-i k \cdot x)
S_{\rho}{}^{\mu}(a_k) \overline S_{\sigma}{}^{\nu}(a^{-1} a_k)
w_{\mu \overline\nu}(a_k^{-1} a a_k) \theta(k^0) \delta(k \cdot k + \mu^2)
\, d^4 k. $$

The positivity condition  (3.15) takes the form
$$(\Phi(f), \Phi(f)) = (2 \pi)^{-3} \int \overline{{\tilde f}^{\rho}}
(k, a_k a) \tilde f^{\sigma}(k, a_k b) \times \tag 3.25 \nopagebreak $$
$$ \times S_{\rho}{}^{\mu}(a^{-1}) \overline S_{\sigma}{}^{\nu}(b^{-1})
w_{\mu \overline\nu}(b a^{-1}) \theta(k^0) \delta(k \cdot k + \mu^2)
\, d^6 a \, d^6 b \, d^4 k  \geq 0,  $$
which is equivalent to the simpler condition
$$ \int \overline{f^{\rho}}(a) f^{\sigma}(b) w_{\rho \overline\sigma}
(b a^{-1}) \, d^6 a \, d^6 b \geq 0. \tag 3.26 $$

The little group $\Cal K$  corresponding to the representative element 
$\hat k$ is defined by the condition$^{21}$
$$ \Lambda(u) \hat k = \hat k, \qquad u \in \Cal K \subset SL(2, \bold C).
\tag 3.27 $$
For $\mu > 0$ we have $\Cal K = SU(2)$, the universal covering of the rotation
group $SO(3)$. For $\mu = 0$, we have $\Cal K = \tilde E(2)$, a double
covering of the Euclidean group $E(2)$.
From eq. (3.16) we get the condition
$$ w_{\rho \overline\sigma}(u a u^{-1}) =
 S_{\rho}{}^{\mu}(u) \overline S_{\sigma}{}^{\nu}(u)
w_{\mu \overline\nu}(a), \qquad  u \in \Cal K. \tag 3.28 $$
From eq.\ (2.11) we also get the property
$$ \overline W_{\rho \overline\sigma}(x, a) =
W_{\sigma \overline\rho}((x, a)^{-1}). \tag 3.29 $$
An equivalent condition is
$$ \overline w_{\rho \overline\sigma}(a) =
w_{\sigma \overline\rho}(a^{-1}), \tag 3.30 $$
which is a consequence of eq.\ (3.26).

Our problem is to find solutions of the conditions (3.26) and (3.28).
Then we have to verify that the product of distributions that appears
in eq.\ (3.24) is meaningful.

If the theory can be interpreted as a theory in Minkowski spacetime,
namely if $V(b) = 1$, the v. e. v.'s satisfy the additional
symmetry property
$$ W_{\rho \overline\sigma}(x, a b^{-1}) = \overline S_{\sigma}{}^{\nu}(b)
W_{\rho \overline\nu}(x, a). \tag 3.31 $$
An equivalent condition is to require that the function
$w_{\rho \overline\sigma}(a)$ defined by eq. (3.23) does not depend on $a$.

\bigpagebreak

\subheading {IV. Local Commutativity}

The study of the distribution $W_{\rho \overline\sigma}(x, a)$ can be
simplified if we find an element $b$ in such a way that
$$ a = b u_3(\alpha) a_3(\beta) b^{-1}, \qquad -2 \pi < \alpha \leq 2 \pi,
\qquad  \beta \geq 0.  \tag 4.1 $$
This is possible outside the four-dimensional submanifold of
$SL(2, \bold C)$ where the eigenvalues of $a$ are equal. Since these
eigenvalues are given by $\exp(\pm \tfrac 1 2 (\beta - i \alpha))$,
we have to avoid the ``singular'' points $\alpha = 0, 2\pi\,$, $\beta = 0$,
namely we have to impose the condition
$$\cosh\beta - \cos\alpha = 2|\sinh(\tfrac 1 2 (\beta -i\alpha)|^2 > 0.
\tag 4.2 $$
We also introduce the quantities
$$ \rho^2 = (\cosh\beta - \cos\alpha)^{-1}
(\cosh\beta\,\, x \cdot x - x \cdot \Lambda(a) x), \tag 4.3 $$
$$ \sigma^2 = (\cosh\beta - \cos\alpha)^{-1}
(\cos\alpha\,\, x \cdot x - x \cdot \Lambda(a) x), \tag 4.4 $$
which, when $a = u_3(\alpha) a_3(\beta)$, take the form
$$ \rho^2 =  (x^1)^2 + (x^2)^2, \qquad
\sigma^2 =  (x^0)^2 - (x^3)^2. \tag 4.5 $$

In order to simplify the formalism, we assume that
$W_{\rho \overline\sigma}(x, a)$
is a tempered distribution in $x$ that depends continuously on $a$. This
is true for the massive free fields described in the next Section.
From eqs.\ (3.9) and (4.1), we see that in the open set defined by eq. (4.2)
we have
$$ W_{\rho \overline\sigma}(x, a) = S_{\rho}{}^{\mu}(b)
\overline S_{\sigma}{}^{\nu}(b)
W_{\mu \overline\nu}(\Lambda(b^{-1}) x, u_3(\alpha) a_3(\beta)). \tag 4.6 $$
The spectral condition implies, as in the Minkowskian field theory, that the
distribution $W_{\rho \overline\sigma}(x, u_3(\alpha) a_3(\beta))$
for fixed values of $\alpha$ and $\beta$  is the boundary value of an
analytic function of $x$ defined in the tube $\text{Im}\, x \in V_+$,
where $V_+$ is the open future cone. From eq.\ (3.9) we also obtain
$$ W_{\rho \overline\sigma}
(\Lambda(u_3(\psi) a_3(\xi))x, u_3(\alpha) a_3(\beta)) =
\tag 4.7  \nopagebreak $$
$$= S_{\rho}{}^{\mu}(u_3(\psi) a_3(\xi))
\overline S_{\sigma}{}^{\nu}(u_3(\psi) a_3(\xi))
W_{\mu \overline\nu}(x, u_3(\alpha) a_3(\beta)). $$
If we fix the variables $\alpha, \beta$,
it is a simple application of the Bargmann Hall Wightman theorem$^{19, 20}$ 
to find by means of eq. (4.7) an analytic continuation of the v. e. v.
(Wightman function)
which is covariant with respect to the complex two-dimensional Lorentz
group acting on the coordinates $x^0, x^3$ and to the real rotations
acting on the coordiantes $x^1, x^2$. The real points which satisfy
the condition $\sigma^2 < 0$ belong to the analyticity domain.

The universal covering of the proper complex Lorentz group is
$SL(2, \bold C) \times SL(2, \bold C)$. We indicate its elements by the
notation $(a, b)$; the real Lorentz transformations correspond to the elements
of the kind $(a, \overline a)$.
If $S(a)$ is the irreducible spinor representation
 $S^{(s, s')}(a) = S^{(s, 0)}(a) \otimes S^{(s', 0)}(\overline a)$, its
analytic continuation is given by
$S^{(s, s')}(a, b) = S^{(s, 0)}(a) \otimes S^{(s', 0)}(b)$.
$\overline S(a)$ is equivalent to $S^{(s', s)}(a)$ and its analytic
continuations is equivalent to $S^{(s', s)}(a, b)$.
We consider the following product of a complex Lorentz transformation
acting on $x^0, x^3$ and a real rotation acting on $x^1, x^2$:
$$ (\exp(- \tfrac 1 2 i \pi \sigma_3), \exp(- \tfrac 1 2 i \pi \sigma_3))
(\exp(- \tfrac 1 2 i \pi \sigma_3), \exp(\tfrac 1 2 i \pi \sigma_3))
= (-1, 1), \tag 4.8 $$
$$ \Lambda(-1, 1) = -1, \qquad S^{(s, s')}(-1, 1) = (-1)^s. \tag 4.9 $$
Then, from the covariance property of the Wightman function we obtain
$$ W_{\rho \overline\sigma}(- x, u_3(\alpha) a_3(\beta)) =
(-1)^{s+s'}  W_{\rho \overline\sigma}(x, u_3(\alpha) a_3(\beta)) \tag 4.10 $$
and the same equality holds for real $x$ if $\sigma^2 < 0$.
The same result is valid if $S$ is a direct sum of irreducible spinor
representations all with the same value of  $(-1)^{s+s'}$.
If we use eq\. (4.6) and the general expression (4.4) for $\sigma^2$,
we see that
$$ W_{\rho \overline\sigma}(- x, a) =
(-1)^{s+s'}  W_{\rho \overline\sigma}(x, a) \tag 4.11 $$
in the open set $\Cal C \subset \tilde{\Cal P}$ defined by the condition
$$ \cos\alpha\,\, x \cdot x - x \cdot \Lambda(a) x < 0. \tag 4.12 $$
One can show that this condition implies eq.\ (4.2).

If we consider a scalar field and we introduce the variables
$$ x = (\epsilon \sigma \cosh\xi', \rho \cos\psi', \rho \sin\psi',
   \epsilon \sigma \sinh\xi'), \qquad \sigma >0,
\tag 4.13 \nopagebreak $$
$$ x = (\epsilon |\sigma| \sinh\xi', \rho \cos\psi', \rho \sin\psi',
\epsilon |\sigma| \cosh\xi'), \qquad \sigma^2 <0, \qquad \epsilon = 
\pm 1,  $$
eq.\ (4.7) shows that $W(x, u_3(\alpha) a_3(\beta))$ does not depend on
the variables $\psi'$ and $\xi'$. From eq.\ (4.10) we see that if
$\sigma^2 < 0$ it does not even depend on $\epsilon$. In conclusion, for
$\sigma^2 < 0$, $W$ can be considered as a distribution
in the variables $\alpha, \beta, x^1, x^2, |\sigma|$ invariant with respect 
to rotations acting on $x^1, x^2$. In particular, since
$a$ and $a^{-1}$ have the same eigenvalues, from eqs.\ (4.3) and (4.4)
we obtain
$$ W(x, a) = W(-x, a) = W(-x, a^{-1}), \qquad 
(x, a) \in \Cal C. \tag 4.14 $$

The commutator or the anticommutator of a free field is a numerical
distribution given by
$$ [\psi_{\rho}(x, a), \psi_{\sigma}^{\dagger}(y, b)]_{\pm} =
W_{\rho \overline\sigma}((x, a)^{-1} (y, b)) \pm
W_{\overline\sigma \rho}((y, b)^{-1} (x, a)) =
\tag 4.15 \nopagebreak $$
$$ = W_{\rho \overline\sigma}((x, a)^{-1} (y, b)) \pm
\overline W_{\overline\rho \sigma}((x, a)^{-1} (y, b)).  $$
If we consider a Hermitean scalar field, and we use eq.\ (3.9),
we have the simpler relation
$$ [\psi(x, a), \psi(y, b)]_- = W(x-y, b a^{-1}) - W(y-x, a b^{-1}) =
\tag 4.16 \nopagebreak $$
$$ = 2 i \text{Im} W((x-y, b a^{-1})  $$
and from eq. (4.14) we get
$$ [\psi(x, a), \psi(y, b)]_- = 0 \qquad \text{for} \quad
(x, a)^{-1} (y, b) \in \Cal C. \tag 4.17 $$
We see that in the formulation of local commutativity, eq.\ (4.12) is
the analog of the inequality $x \cdot x > 0$ in Minkowski field theory.

In the general case, if we impose a local (anti)commutativity condition
of the kind
$$ [\psi_{\rho}(x, a), \psi_{\sigma}^{\dagger}(y, b)]_{\pm} = 0
\qquad \text{for} \quad (x, a)^{-1} (y, b) \in \Cal C,  \tag 4.18 $$
from eqs. (4.11) and (4.15) we get
$$ W_{\overline\rho \sigma}(x, a) = \mp (-1)^{s+s'}
\overline W_{\rho \overline\sigma}(-\overline x, a), \qquad
(x, a) \in \Cal C.   \tag 4.19 $$
This relation can be continued analytically in the tube
$\text{Im}\, x \in V_+$ and, as an equality of distributions, it holds for
any real $x$. It is compatible with the spectral and covariance conditions,
but it satisfies the positivity condition only if $\mp (-1)^{s+s'} =1$.
In conclusion, the local (anti)commutativity relation (4.18) is
satisfied if we put
$$ W_{\overline\rho \sigma}(x, a) =
\overline W_{\rho \overline\sigma}(-x, a),  \tag 4.20 $$
and the statistics is determined by $(-1)^{s+s'}$.
We shall see that $s+s'$ is not directly related to the spin of the
particles described by the field and the usual relation between spin
and statistics is not necessarily valid. From eq.\ (3.24) we see
that eq.\ (4.20) is equivalent to the condition
$$ w_{\overline\rho \sigma}(a) =
\overline w_{\rho \overline\sigma}(a).  \tag 4.21 $$
For a Hermitean field this condition requires that $w_{\rho \sigma}(a)$
is real.

It is interesting to study the integral (3.24) in the scalar case with
more detail. If $w(a) = 1$, eq.\ (3.24) gives the usual v. e. v.
for the Minkowskian free scalar field, namely
$$ W(x) = i (4 \pi)^{-1} \epsilon(x^0) \delta(s^2)  +
(2 \pi)^{-2} \theta(-s^2)  \mu |s|^{-1} K_1(\mu |s|) +
\tag 4.22 \nopagebreak  $$
$$ + (8 \pi)^{-1} \theta(s^2)  \mu s^{-1}
\left(Y_1(\mu s) - i \epsilon(x^0) J_1(\mu s)\right), \qquad
s^2 = -x \cdot x = \sigma^2 - \rho^2,  $$
where $\epsilon(x^0)$ is the sign of $x^0$. The corresponding commutator
vanishes for $\sigma^2 < \rho^2$.

If the function $w(a_k^{-1} a a_k)$ decreases for large $k$
the distribution $W(x, a)$ is less singular.  In order to simplify
the integral (3.24), it is convenient to write the four-vector $k$
in the following way
$$ k =  (q \cosh\xi, p \cos\psi, p \sin\psi, q \sinh\xi), \qquad
q = (p^2 + \mu^2)^{\tfrac 1 2 }, \tag 4.23 \nopagebreak $$
$$ 0 \leq p < \infty, \quad -\infty < \xi < \infty, \quad
0 \leq \psi < 2 \pi, \nopagebreak $$
$$ (2 k^0)^{-1} \, d^3 \bold k =
\tfrac 1 2 \, p dp \, d \xi \, d \psi  $$
and to choose
$$ a_k = u_3(\psi) a_3(\xi) a_{\eta}. \tag 4.24 $$
For $\mu > 0$ we take
$$ a_{\eta} = a_1(\eta), \qquad
p = \mu \sinh\eta, \qquad q = \mu \cosh\eta, \qquad
0 \leq \eta < \infty \tag 4.25 $$
and for $\mu = 0$
$$ a_{\eta} = a_1(\eta) u_2(\tfrac 1 2 \pi), \qquad
p = q = \exp\eta, \qquad
- \infty < \eta < \infty. \tag 4.26 $$
From eq.\ (4.13) we get
$$ k \cdot x =  - \epsilon q \sigma \cosh(\xi - \xi')
+ p \rho \cos(\psi - \psi'), \qquad \sigma^2 > 0, \tag 4.27 \nopagebreak $$
$$ k \cdot x =  \epsilon q |\sigma| \sinh(\xi - \xi')
+ p \rho \cos(\psi - \psi'), \qquad \sigma^2 < 0. $$

The integrations over the variables $\xi$ and $\psi$ in eq\. (3.24) can be
performed in terms of Bessel functions$^{27}$ and we get the following formula
$$ W(x, a) = (2 \pi)^{-1} \int_0^{\infty}
\Delta(q \sigma, \epsilon(x^0))
 J_0(p \rho) w(\alpha, \beta, p) p \, d p, \tag 4.28 $$
where
$$ \Delta(q \sigma, \epsilon(x^0)) =
(2 \pi)^{-1} \theta(- \sigma^2) K_0(q |\sigma|)
- \tfrac 1 4 \theta(\sigma^2) \left(Y_0(q \sigma) -
 i \epsilon(x^0) J_0(q \sigma)\right) \tag 4.29   $$
and
$$ w(\alpha, \beta, p) = w(a_{\eta}^{-1} u_3(\alpha) a_3(\beta) a_{\eta}).
\tag 4.30 $$

The commutator of a scalar Hermitean field follows from eqs.\ (4.16) and
(4.28) and has the form
$$ 2 i \text{Im} W(x, a) = i (4 \pi)^{-1} \theta(\sigma^2) \epsilon(x^0)
 \int_0^{\infty} J_0(q \sigma)  J_0(p \rho) w(\alpha, \beta, p) p \, dp.
\tag 4.31 $$
We see that it vanishes for $\sigma^2 < 0$.
Eq.\ (4.31) can be considered as a Hankel transformation$^{27}$.
By considering the corresponding inverse transformation, one
can easily see that a necessary condition to have a commutator vanishing for
large $\rho$ is that $w(\alpha, \beta, p)$ is an entire analytic function
of $p^2$. This does not happen for the elementary fields considered in
the following Sections.

When $w(\alpha, \beta, p)$ is an even function of $p$,
it is useful to rewrite eq.\ (4.31) as an integral in the complex $p$ plane
$$ 2 i \text{Im} W(x, a) = i 2^{-3} \pi^{-1} \theta(\sigma^2) \epsilon(x^0)
 \int_C J_0(q \sigma) H^{(1)}_0(p \rho)
w(\alpha, \beta, p) p \, dp, \tag 4.32 $$
where the integration path $C$ lies just above the real axis.
For $|p| \to \infty$, $\text{Im}\,p >0$, we have
$$ p J_0(q \sigma) H^{(1)}_0(p \rho) \simeq \pi^{-1}
(\rho \sigma)^{-\tfrac 1 2} \exp(ip(\rho - \sigma)). \tag 4.33 $$
If $w=1$ and $\rho > \sigma$, we can close the integration path
at infinity in the upper half plane and the integral
vanishes in accord with eq.\  (4.22). Actually, the integral (4.32)
is meaningful only in the sense of distribution theory; in order to deal
with a convergent integral, one can multiply the integrand by $(p + i)^{-\nu}$
and take the limit $\nu \to 0$ at the end. If $w$ has singularities in the
upper half plane, one has to take into account their contributions.

\bigpagebreak

\subheading {V. Positive-Mass Free Fields}

In the positive-mass case, we can find the function
$w_{\rho \overline\sigma}(a)$ by exploiting
the positivity properties of the matrix elements $D^{M c}_{jmj'm'}(a)$
of the i. u. r.'s of $SL(2, \bold C)$ described in refs.\ 22-26.
The parameter $c$ is the same as in ref.\ 22 and is called
$\lambda$ in ref.\ 24. The parameter $M$ is the same as in ref.\ 24
and corresponds to the parameter $-\tfrac 1 2 m = \pm k_0$ of ref.\ 22.
For the i. u. r.'s of the principal series, $c$ lies on the imaginary axis
and $M$ is an integral or half-integral number.  For the i. u. r.'s of the
supplementary series, $M=0$ and $-1 < c < 1$.
The representations $D^{M c}$ and $D^{-M, -c}$ are unitarily equivalent.
The indices $j, j'$ take the values $|M|, |M|+1, \ldots$ and,
if we indicate by $R^j_{mm'}(u)$ the $(2j+1)$-dimensional representation
of $SU(2)$, we have
$$ D^{M c}_{jmj'm'}(u) = \delta_{jj'} R^j_{mm'}(u), \qquad u \in SU(2).
\tag 5.1 $$

Then we put
$$ w_{\rho \overline\sigma}(a) = (2s+1) (2J+1)^{-1}
\sum_{m n n'} C_{js}(J, m; n', \rho)
C_{js}(J, m; n, \sigma) D^{M c}_{jnjn'}(a), \tag 5.2 $$
where $C$ indicates the Clebsch-Gordan coefficients. The parameters
$M, c, j, J, s$ characterize the theory; they are fixed and no sum over them
is understood. After some calculation we have
$$ w_{\rho \overline\sigma}(u a u^{-1}) = R^s_{\rho \mu}(u)
\overline R^s_{\sigma \nu}(u) w_{\mu \overline\nu}(a) \tag 5.3 $$
and we see that eq.\ (3.28) is satisfied if
$$ S_{\rho}{}^{\sigma}(u) = R^s_{\rho \sigma}(u), \qquad
u \in SU(2). \tag 5.4 $$
If we indicate by $S^{(s, s')}$ an irreducible spinor representation,
we can put, with an appropriate choice of the basis,
$$ S(a) = S^{(s, 0)}(a) \quad \text{or} \quad
S(a) =  S^{(0, s)}(a). \tag 5.5 $$

In order to prove the positivity property, we substitute eq.\ (5.2)
into eq.\ (3.25) and we get the positive expression
$$ (\Phi(f), \Phi(f)) = \int \sum_{mj'm'} |f_{mj'm'}(k)|^2
\theta(k^0) \delta(k \cdot k + \mu^2) \, d^4 k, \tag 5.6 $$
where
$$ f_{mj'm'}(k) = (2 \pi)^{-\tfrac 3 2}
(2s+1)^{\tfrac 1 2} (2J+1)^{-\tfrac 1 2} \times \tag 5.7 \nopagebreak $$
$$ \times C_{js}(J, m; n, \mu)
\int \tilde f^{\rho}(k, a_k a) \overline S_{\rho}{}^{\mu}(a^{-1})
D^{M c}_{jnj'm'}(a) \, d^6 a.  $$
It is clear that the quantity (5.7) is the wave function in momentum space of
the one-particle state $\Phi(f)$ and that its indices represent the spin
and the internal quantum numbers.

Eq\. (2.15) takes the form
$$ U^{(1)}(y, b) \Phi(f) = \Phi(f'), \qquad f^{\prime\rho}(x, a) =
f^{\rho}(\Lambda(b^{-1})(x - y), b^{-1} a). \tag 5.8 $$
The Fourier transformation (3.14) gives
$$ \tilde f^{\prime\rho}(k, a) =
\exp(-i k \cdot y) \tilde f^{\rho}(k', b^{-1} a), \qquad
k' = \Lambda(b^{-1}) k   \tag 5.9 $$
and from eq\. (5.7) we obtain
$$f'_{mj'm'}(k) = \exp(-i k \cdot y) R^J_{mm''}(a_k^{-1} b a_{k'})
f_{m''j'm'}(k'), \tag 5.10  \nopagebreak $$
$$ a_k^{-1} b a_{k'} \in SU(2).  $$
We see that the wave function transforms according to the i. u. r. with
mass $\mu > 0$ and spin $J$ defined by Wigner$^{21}$. Eq. (5.7) shows that
spin has a double origin, namely the dependence of $f^{\rho}(x, a)$ on the
index $\rho$ and on the group element $a$.  From eqs.\ (3.24) and (5.2)
we have
$$ W_{\rho \sigma}(x, -a) = (-1)^{2J} W_{\rho \sigma}(x, a) \tag 5.11 $$
and a similar formula holds for the field $\psi_{\rho}(x, a)$, which
is a one or two-valued function on $\Cal P$ if $2J$ is, respectively, even
or odd.

Eq.\ (2.16) takes the form
$$V^{(1)}(b) \Phi(f) = \Phi(f'), \qquad
f^{\prime\rho}(x, a) = f^{\sigma}(x, ab) \overline 
S_{\sigma}{}^{\rho}(b^{-1}),  \tag 5.12  $$
and from eqs.\ (3.14) and (5.7) we get
$$ f'_{mj'm'}(k) = f_{mj''m''}(k)  D^{M c}_{j''m''j'm'}(b^{-1}) =
\overline D^{M c}_{j'm'j''m''}(b) f_{mj''m''}(k). \tag 5.13 $$
We see that the wave function transforms according to the i. u. r.
$\overline D^{M c}$ of the internal symmetry group $SL(2, \bold C)$.
This i. u. r. is equivalent to  $D^{M c}$.

The general formalism described above becomes simpler in some special cases.
If $s = 0$, $S(a) = 1$, we have a scalar field.
It is $J = j$ and eq.\ (5.2) takes the form
$$ w(a) = w^{Mcj}(a) = (2j + 1)^{-1} \sum_m D^{M c}_{jmjm}(a). \tag 5.14 $$
If $j = M = 0$, we have $J = s$ and
$$ w_{\rho \overline\sigma}(a) = \delta_{\rho \sigma} D^{0 c}_{0000}(a) =
\delta_{\rho \sigma} w^{0c0}(a). \tag 5.15 $$

In order to give more explicit expressions for the functions $w^{Mcj}(a)$,
we write the i. u. r.'s  of $SU(2)$ and of $SL(2, \bold C)$ in the form
$$ R^j_{mm'}(u_3(\phi) u_2(\theta) u_3(\psi)) =
\exp(-i m \phi) r^j_{mm'}(\theta) \exp(-i m' \psi), \tag 5.16 $$
$$ D^{M c}_{jmj'm'}(u a_3(\zeta) u') =
\sum_{m''} R^j_{mm''}(u) d^{M c}_{m''jj'}(\zeta) R^{j'}_{m''m'}(u').
\tag 5.17 $$
Then the  function defined by eq\. (5.14) can be written as
$$ w^{M c j}(u a_3(\zeta) u') =
w^{M c j}(a_3(\zeta) u_3(\phi + \psi) u_2(\theta) ) =
      \tag 5.18  \nopagebreak  $$
$$ = (2j + 1)^{-1}
\sum_m d^{M c}_{mjj}(\zeta) r^j_{mm}(\theta) \exp(-im (\phi +\psi)),
\qquad u'u = u_3(\phi) u_2(\theta) u_3(\psi). $$
The quantities $d^{M c}_{mjj}(\zeta)$ are given in refs.\ 23, 24
in terms of elementary functions. It follows that the expression
(5.18) too is a combination of elementary functions.
For instance, we have
$$ w^{0 c 0}(u a_3(\zeta) u') =
d^{0 c}_{000}(\zeta) = (c \sinh(\zeta))^{-1} \sinh(c \zeta), \tag 5.19  $$
but the expressions become more and more complicated when the parameters
$M$ and $j$ increase.  An useful integral representation of $w^{M c j}(a)$
is given in the Appendix A, where one derives also a simple approximate
expression valid for $a \to 1$.

Note that we have
$$ \lim_{c \to 1} w^{0 c 0}(a) =
\lim_{c \to 1} D^{0 c}_{0000}(a) = 1. \tag 5.20 $$
In this limit the distribution $w_{\rho \overline\sigma}(a)$ given by eq\.
(5.15) becomes independent of $a$ and we have a theory which can be defined
in Minkowski spacetime. From the unitarity condition we also have
$$ \lim_{c \to 1} D^{0 c}_{00jm}(a) = \lim_{c \to 1} D^{0 c}_{jm00}(a) = 0,
\qquad j>0, \tag 5.21 $$
and if $j = M = 0$ the components of the wave function (5.7) tend to zero
for  $j' > 0$.
We conclude that the theories with $M = j = 0$ can possibly describe
small deviations from the known physical theories. For other
values of the parameters, we get theories which can only apply to
the unknown physics of masses and energies beyond the Planck mass.
From the mathematical point of view,
eqs.\ (5.20) and (5.21) show that in the limit $c \to 1$, the
representation $D^{0c}$ becomes reducible and we have$^{22}$
$$ \lim_{c \to 1} D^{0 c} = 1 \oplus D^{1, 0}. \tag 5.22 $$.

We see from eq.\ (3.18) that $\tilde W_{\rho \overline\sigma}(k, a)$
can be considered as a tempered distribution in the variables $k$ that
depends continuously on the group element $a$ and a similar statement holds
for its Fourier transform  $W(x, a)$.
If $a = 1$, eq.\ (5.2) gives
$$ w_{\rho \overline\sigma}(1) =  \delta_{\rho \sigma}
 \tag 5.23 $$
and from eq\. (3.24) we see that
$W_{\rho \overline\sigma}(x, 1)$ is the v. e. v. of a
field in Minkowski space, given, in the scalar case, by eq.\ (4.22).
If $a \neq 1$, the function $w_{\rho \overline\sigma}(a_k^{-1} a a_k)$
decreases for large $k$ and the distribution $W_{\rho \overline\sigma}(x, a)$
is less singular.

For a scalar field we can use eq.\ (4.28) and from eq. (4.30) we obtain
$$ w(\alpha, \beta, p) = w(a_3(\zeta) u_3(\phi) u_2(\theta)),
\tag 5.24 $$
where
$$ \cosh\zeta = (\cosh\eta)^2 \cosh\beta - (\sinh\eta)^2 \cos\alpha =
 \tag 5.25 \nopagebreak $$
$$ = \cosh\beta + p^2 \mu^{-2} (\cosh\beta - \cos\alpha), $$
$$ \cos\frac \theta 2 \exp(i\frac \phi 2) =
\left(\cosh \frac \zeta 2 \right)^{-1}
\cos \frac \alpha 2 \cosh \frac \beta 2
+i \left(\sinh \frac \zeta 2 \right)^{-1}
\sin \frac \alpha 2 \sinh \frac \beta 2. \tag 5.26 $$
For $j = M = 0$, we see that the expression (5.19) is an analytic function
of $\cosh\zeta$ with a branch point at $\cosh\zeta = -1$. From eq.\
(5.25) we see that it is not an entire analytic function of $p^2$ and the
commutator function (4.31) cannot vanish for large $\rho$.

\bigpagebreak

\subheading {VI. Zero-Mass Free Fields}

In order to use the method described in the preceding Section in the
zero-mass case, we have to write the i. u. r.'s of $SL(2, \bold C)$
in a basis that evidentiates the decomposition of the representation space
into spaces where i. u. r.'s of $\tilde E(2)$ operate. In this case, we have
a direct integral decomposition and we have to introduce a ``continuous''
basis$^{28}$.

We start from the realization of the operator $D^{M c}(a)$ in a space of
functions of a complex variable $z$:
$$ [D^{M c}(a) f](z) = \tag 6.1 \nopagebreak $$
$$ = (a_{21} z +a_{11})^{-c+M-1} (\overline{a_{21} z +a_{11}})^{-c-M-1}
f\left((a_{22} z +a_{12})(a_{21} z +a_{11})^{-1}\right), $$
where
$$ a = \left( \matrix a_{11} & a_{12} \\ a_{21} & a_{22} \endmatrix \right)
\in SL(2, \bold C). \tag 6.2 $$
This representation is equivalent to the one defined in refs.\ 22, 23,
that in our notation is given by $D^{-M, -c}(u_1(\pi) a u_1(-\pi))$.

The elements of the little group $\tilde E(2)$ have the form
$$ h = \left(\matrix \exp(-\tfrac 1 2 i \phi) & 
\xi \exp(-\tfrac 1 2 i \phi) \\
0 & \exp(\tfrac 1 2 i\phi) \endmatrix \right) \in \tilde E(2), \tag 6.3 $$
with $\phi$ real and $\xi$ complex. We have
$$ [D^{M c}(h) f](z) = \exp(-i M \phi)
f(\exp(i \phi) z + \xi). \tag 6.4 $$
We see that $\tilde E(2)$ acts on the complex $z$ plane by means of Euclidean
transformations. If we introduce the (non normalizable) basis vectors
$$ f_{\kappa m}(z) = (z |z|^{-1})^{-m} J_{-m}(\kappa |z|), \qquad \kappa > 0,
\qquad m = 0, \pm 1, \pm 2, \ldots, \tag 6.5 $$
the matrix that represents an Euclidean transformation is diagonal
in the index $\kappa$, namely we have
$$ f_{\kappa m}(\exp(i \phi) z + \xi) = f_{\kappa m'}(z) R^{\kappa}_{m'm}(h),
 \tag 6.6 $$
where the matrix
$$ R^{\kappa}_{m'm}(h) = \exp(-i m' \phi) (\xi |\xi|^{-1})^{m'-m}
J_{m'-m}(\kappa |\xi|) \tag 6.7 $$
is unitary$^{26}$.  From eq.\ (6.4), we get
$$ [D^{M c}(h) f_{\kappa m}](z) =  \exp(-i M \phi) R^{\kappa}_{m'm}(h)
f_{\kappa m'}(z). \tag 6.8 $$
We see from eq.\ (6.7) that the factor $\exp(-i M \phi)$ can be eliminated
by means of a ``translation'' of the indices $m', m$, that is irrelevant
if $M$ is an integer.  If $M$ is half-odd, we get a representation of
$\tilde E(2)$ that is double-valued on $E(2)$.

In analogy with eq.\ (5.14) we put
$$ w(a) = \sum_m (f_{\kappa m}, D^{M c}(a) f_{\kappa m}). \tag 6.9 $$
Also in this case, the conditions (3.26) and (3.28) are satisfied.
In the same way as in the preceding Section, we can show that
the fields constructed by means of the distributions found above represent
zero-mass particles with an infinite helicity spectrum$^{21}$ and with
internal quantum numbers described by the i. u. r. $D^{M c}$ of
$SL(2, \bold C)$. Since particles of this kind are not observed in nature,
we shall not discuss these fields with more detail.

The zero-mass particles present in nature
have only one value $m$ of the helicity (two if parity is taken into account)
and they are described by one-dimensional i. u. r.'s of the
little group$^{21}$, of the kind
$$ R^m(h) = \exp(-i m \phi), \qquad h \in \tilde E(2), \qquad
m = 0, \pm \tfrac 1 2 , \pm 1, \ldots \tag 6.10 $$
where $h$ is given by eq\. (6.3). These i.u.r.'s are not contained in the
i. u. r.'s of $SL(2, \bold C)$ and we have to use a different
method in order to find the corresponding $w$ functions.
We propose the following solutions, without explaining how they have been
obtained:
$$ w_0^{0c}(a) = |a_{21}|^{2c-2}, \qquad 0 < c <1, \tag 6.11 $$
$$ w_0^{Mc}(a) =   \delta^2(a_{21}) a_{22}{}^{c-M} \overline a_{22}{}^{c+M},
\qquad \text{Re}\, c = 0. \tag 6.12 $$
It is easy to show that the symmetry condition (3.28) is satisfied.
The positivity condition (3.26) follows from the formulae
$$ w_0^{0c}(b a^{-1}) = \tag 6.13 \nopagebreak $$
$$ = \int |a_{22}|^{2c} \delta^2(a_{21} - z'a_{22}) |z' - z|^{2c-2}
|b_{22}|^{2c} \delta^2(b_{21} - z b_{22}) \, d^2z \, d^2z', $$
$$ w_0^{Mc}(b a^{-1}) = \tag 6.14 \nopagebreak $$
$$ = \int \overline a_{22}{}^{-c-M} a_{22}{}^{-c+M} \delta^2(a_{21} - za_{22})
b_{22}{}^{c-M} \overline b_{22}{}^{c+M} \delta^2(b_{21} - z b_{22}) 
\, d^2z. $$
Note that $|z' - z|^{2c-2}$ is a positive integral kernel, the one that
defines the scalar product in the space of an i. u. r. of the
supplementary series$^{22, 23}$.

The norm (3.25) in the case (6.11) can be written as
$$ (\Phi(f), \Phi(f)) = \int \overline f(k, z') |z' - z|^{2c-2} f(k, z)
\theta(k^0) \delta(k \cdot k) \, d^2 z \, d^2 z' \, d^4 k, \tag 6.15 $$
and in the case (6.12) we get
$$ (\Phi(f), \Phi(f)) = \int |f(k, z)|^2
\theta(k^0) \delta(k \cdot k) \, d^2 z \, d^4 k, \tag 6.16 $$
where in both cases
$$ f(k, z) = (2 \pi)^{-\tfrac 3 2} \int \tilde f(k, a_k a)
a_{22}{}^{c-M} \overline a_{22}{}^{c+M}
\delta^2(a_{21} - z a_{22}) \, d^6 a. \tag 6.17 $$
Of course, in the first case we have to put $M = 0$.

If we consider a Poincar\'e transformation of the kind (5.8), (5.9), the wave
function (6.17) transforms in the following way
$$ f'(k, z) =  \exp(-i k \cdot y) R^M(a_k^{-1} b a_{k'})
f(k', z), \tag 6.18 $$
where $k'$ is given by eq. (5.9).
This is the transformation property$^{21}$ of the wave function of a
particle of zero mass and helicity $M$.

Under the ``internal'' transformation (5.12) the wave function transforms as
$$ f'(k, z) = (b_{22} + z b_{12})^{-c+M-1}
(\overline{b_{22} + z b_{12}})^{-c-M-1} f(k, z'), \tag 6.19 \nopagebreak $$
$$ z' = (b_{21} + z b_{11})(b_{22} + z b_{12})^{-1}, $$
namely according to the i. u. r. $D^{-M,-c}$ defined in refs.\ 22, 23,
which is equivalent to the i. u. r. $D^{M c}$. Note that, if we fix up to
equivalence the i. u. r. $D^{M c}$, we have two possible theories with
helicity $\pm M$.

In order to compute the integral (3.24), we use the variables introduced
in Sec. IV. If we put
$$ \tilde a = a_{\eta}^{-1} u_3(\alpha) a_3(\beta) a_{\eta}, \tag 6.20 $$
from  eq.\ (4.26) we get
$$ \tilde a_{21} = - p \sinh(\tfrac 1 2 (\beta - i \alpha)), \qquad
\tilde a_{22} =  \cosh(\tfrac 1 2 (\beta - i \alpha)). \tag 6.21 $$

We see that the quantity $\delta^2(\tilde a_{21})$ that appears in eq.\ (6.12)
is rather badly defined when considered as a distribution in the variables
$k^1, k^2$ for fixed values of $\alpha, \beta$ satisfying eq.\ (4.2).
Other difficulties arise when on tries to perform the integrals
(3.24) or (4.28). We conclude that,
even if the expression (6.12) satisfies the required positivity and
symmetry conditions,  it does not give rise to a well-behaved field theory.
All the calculations based on this expression have a purely formal character.

When we substitute eqs.\ (6.11) and (6.21) into eqs.\ (4.28) and (4.31),
the integrals can be perfomed in terms of Legendre functions$^{27, 29}$
and we get
$$ W^{0c}_0(x, a) = 2^{c-3} \pi^{-2} (\Gamma(c))^2
(\cosh\beta - \cos\alpha)^{c - 1} \times  \tag 6.22 \nopagebreak $$
$$ \times (\rho^2 - \sigma^2)^{-c}
P_{c-1}((\sigma^2 + \rho^2)(\sigma^2 - \rho^2)^{-1}), \qquad \sigma^2 < 0. $$
$$ 2i \text{Im}\, W^{0c}_0(x, a) = i 2^{c-2} \pi^{-1} \epsilon(x^0)
\theta(\sigma^2) \Gamma(c) (\Gamma(1-c))^{-1} \times \tag 6.23 \nopagebreak $$
$$ \times (\cosh\beta - \cos\alpha)^{c - 1} |\sigma^2 - \rho^2|^{-c}
P_{c-1}((\sigma^2 + \rho^2)|\sigma^2 - \rho^2|^{-1}).   $$
The real part for $\sigma^2 > 0$ can be obtained by analytic continuation
of eq.\ (6.22). In the special case $\rho = 0$, we have
$$ W^{0c}_0(x, a) =  2^{c-3} \pi^{-2} (\Gamma(c))^2
(\cosh\beta - \cos\alpha)^{c - 1} \times  \tag 6.24 \nopagebreak $$
$$\times |\sigma^2|^{-c} \left(\theta(-\sigma^2) +
\theta(\sigma^2) \exp(i \pi c \epsilon(x^0)) \right). $$

Since we are not able to build fields starting from eq.\ (6.12), in order
to describe particles with a non-vanishing helicity we have to use
non-scalar fields. From eq. (6.3) we have
$$ h \left( \matrix 1 \\ 0 \endmatrix \right) = \exp(- \tfrac 1 2 i \phi)
\left( \matrix 1 \\ 0 \endmatrix \right). \tag 6.25 $$
We remark that the spinor representations $S^{(s, 0)}(h)$ and $S^{(0, s)}(h)$
are equivalent to symmetrized tensor products of $s$ matrices equal,
respectively, to $h$ or $\overline h$ and we adopt conventions in
agreement with eqs.\ (5.4) and (5.5). Then, if we put
$$  S(a) = S^{(m, 0)}(a), \qquad m \geq 0,
\tag 6.26 \nopagebreak $$
$$  S(a) = S^{(0, |m|)}(a), \qquad m \leq 0, $$
we have
$$  S_{\rho}{}^m(h) = \exp(-im \phi) \delta_{\rho}^m, \qquad
 h \in \tilde E(2). \tag 6.27 $$
As a consequence, the expression
$$ w_{\rho \overline\sigma}(a) = \delta_{\rho m} \delta_{\sigma m}
|a_{21}|^{2c-2} \tag 6.28 $$
satisfies the symmetry condition (3.28).

Eq.\ (6.15) is still valid if we modify eq.\ (6.17) in the following way
$$ f(k, z) = (2 \pi)^{-\tfrac 3 2} \int \tilde f^{\rho}(k, a_k a)
\overline S_{\rho}{}^m(a^{-1})  |a_{22}|^{2c}
\delta^2(a_{21} - z a_{22}) \, d^6 a. \tag 6.29 $$
Eq.\ (6.18) holds with $M$ replaced by $m$ and eq.\ (6.19) holds with
$M = 0$; this means that the theory describes zero masss particles with
helicity $m$ that transform according to the representation $D^{0c}$
of the internal symmetry group.  In the limit $c \to 1$, the expression
(6.28) becomes independent of $a$ and we get a Minkowskian theory
of the kind usually adopted to describe neutrinos, photons or gravitons.

\bigpagebreak

\subheading {VII. Field Equations}

Now we find the wave equations satisfied by the scalar fields
defined in the preceding Sections; non-scalar fields will be treated 
elsewhere. We indicate by $ L_{ik} = -  L_{ki} $ the generators of the
right translations on $SL(2,\bold C)$, considered as
differential operators acting on smooth functions defined on the group.
They satisfy the commutation relations
$$ [ L_{ik},  L_{rs}] = g_{ir}  L_{ks} - g_{kr}  L_{is}
- g_{is}  L_{kr} + g_{ks}  L_{ir} \tag 7.1 $$
of $sl(2,\bold C)$, commute
with the left translations and have the following commutation  property
with the finite right translation represented by the operator $T(b)$:
$$ T(b^{-1}) L_{ik} T(b) = \Lambda_i{}^r(b) \Lambda_k{}^s(b) L_{rs}.
\tag 7.2 $$
We also need the generators  $ L'_{ik} $ of the left translations,
which satisfy commutation relations with the opposite sign. We have
$$  L'_{ik} = \Lambda_i{}^r(a) \Lambda_k{}^s(a)  L_{rs} \tag 7.3 $$
and their commutation relation with a finite left translation $T'(b)$ is
$$ T'(b) L'_{ik} T'(b^{-1}) = \Lambda_i{}^r(b) \Lambda_k{}^s(b)  L'_{rs}.
\tag 7.4 $$

We also consider the generators $A_k$ and $A_{ik}$ of the right translations
on the group $\tilde{\Cal P}$, that satisfy the Poincar\'e
Lie algebra. They act on functions of the kind $f(x, a)$ in the following way:
$$ A_i = \Lambda^k{}_i(a) \frac {\partial}{\partial x^k}, \qquad
   A_{ik} =  L_{ik}.  \tag 7.5 $$
For the generators of the left translations of $\tilde{\Cal P}$ we have
$$ A'_i = \frac {\partial}{\partial x^k}, \qquad
A'_{ik} =  L'_{ik} + x_i \frac {\partial}{\partial x^k}
- x_k \frac {\partial}{\partial x^i}.  \tag 7.6 $$
From eqs.\ (3.24), (7.5) and (7.6) we obtain immediately the
differential equation
$$ g^{ik} A_i A_k W(x,a) = g^{ik} A'_i A'_k W(x,a) =
 {\mu}^2  W(x, a), \tag 7.7  $$
which is essentially the Klein Gordon equation.

If the function $w(a)$ is a linear combination of matrix elements of
the representation $D^{Mc}(a)$, as in eqs.\ (5.2) and (6.9), it satisfies
the differential equations$^{22, 23}$
$$ \tfrac 1 2 g^{ir} g^{ks}  L_{ik}  L_{rs} w(a) =
  \tfrac 1 2 g^{ir} g^{ks}  L'_{ik}  L'_{rs} w(a) =
 (1 - c^2 - M^2)  w(a), \tag 7.8  $$
$$ \tfrac 1 8 e^{ikrs}  L_{ik}  L_{rs} w(a) =
   \tfrac 1 8 e^{ikrs}  L'_{ik}  L'_{rs} w(a) =
 -i M c w(a). \tag 7.9  $$

The functions (6.11) and (6.12) are not defined in terms of matrix
elements, but it is easy to show directly that they satisfy the differential
equations
$$  (L'_{10} - L'_{31}) w(a) = 0, \qquad
(L'_{20} + L'_{23}) w(a) = 0, \tag 7.10 $$
$$ L'_{12} w(a) = - i M w(a), \qquad L'_{30} w(a) = (1 - c) w(a). \tag 7.11 $$
By means of the formulas
$$  \tfrac 1 2 g^{ir} g^{ks} L'_{ik} L'_{rs} =  \tag 7.12  \nopagebreak $$
$$ = -(L'_{10} + L'_{31})(L'_{10} - L'_{31})
- (L'_{20} - L'_{23})(L'_{20} + L'_{23}) +
 (L'_{12})^2 - (L'_{30})^2 + 2 L'_{30},  $$
$$ \tfrac 1 8 e^{ikrs} L'_{ik} L'_{rs} = \tag 7.13 \nopagebreak $$
$$ = \tfrac 1 2 (L'_{20} - L'_{23})(L'_{10} - L'_{31})
- \tfrac 1 2 (L'_{10} + L'_{31})(L'_{20} + L'_{23})
- L'_{30} L'_{12} + L'_{12}. $$
we can derive also in this case the eqs.\ (7.8) and (7.9).

The Casimir operators that appear in eqs.\ (7.8)
and (7.9) commute with the left and the right translations on $SL(2, \bold C)$
and from eqs.\ (3.24) and (7.5) we get, for a scalar field,
$$ \tfrac 1 2 g^{ir} g^{ks} A_{ik} A_{rs} W(x,a) =
(1 - c^2 - M^2)  W(x, a), \tag 7.14 $$
$$ \tfrac 1 8 e^{ikrs} A_{ik} A_{rs} W(x, a) = -i M c W(x, a). \tag 7.15 $$

Eqs.\ (7.7), (7.14) and (7.15) identify the mass and the internal quantum
numbers and hold for all the scalar fields. There are other equations
that identify spin or helicity. In the positive-mass case
the generators of the little group $\Cal K = SU(2)$ are
$ L_{12},  L_{23},  L_{31}$, and from eqs.\ (5.1) and (5.2), taking
into account the properties of the representations of $SU(2)$, we get
$$ \left(( L'_{12})^2 + ( L'_{23})^2 +( L'_{31})^2 \right)
   w(a) = - j(j+1) w(a).  \tag 7.16 $$
If $j = 0$, we have the stronger result
$$  L'_{12} w(a) =   L'_{23} w(a) =  L'_{31} w(a) =  0. \tag 7.17 $$
From eqs.\ (3.19) and (7.16) we get
$$ \tfrac 1 4 e^{jirs} \hat k_i L'_{rs} e_j{}^{kpq} \hat k_k L'_{pq}
w(a) = - \mu^2 j(j+1) w(a) \tag 7.18 $$
and from eqs. (3.21), (3.22) and (7.4)
$$ \tfrac 1 4 e^{jirs} k_i L'_{rs} e_j{}^{kpq} k_k L'_{pq}
 w(k, a) = - \mu^2 j(j+1) w(k, a). \tag 7.19 $$
By substitution into eq.\ (3.24) we obtain  the differential equation
$$ g_{ik} \Sigma^{\prime i} \Sigma^{\prime k} W(x, a) =
 \mu^2 j(j+1) W(x, a),  \tag 7.20 $$
where
$$ \Sigma^{\prime i} = \tfrac 1 2 e^{ijrs} A'_j A'_{rs}=
 \tfrac 1 2 e^{ijrs} \frac {\partial}{\partial x^j} L'_{rs}. \tag 7.21 $$
If we also introduce the differential operators
$$ \Sigma^i = \tfrac 1 2 e^{ijrs} A_j A_{rs} = \Lambda_k{}^i(a)
\Sigma^{\prime k},  \tag 7.22 $$
we have
$$ g_{ik} \Sigma^i \Sigma^k W(x, a) = \mu^2 j(j+1) W(x, a).  \tag 7.23 $$
If $j = 0$ we have the stronger result
$$ \Sigma^i  W(x, a) = \Sigma^{\prime i} W(x, a) = 0. \tag 7.24 $$

In the zero-mass case, the generators of the little group
$\Cal K = \tilde E(2)$ are $L_{12}, (L_{10} - L_{31}), (L_{20} + L_{23})$.
From eqs.\ (6.8) and (6.9) we obtain the differential equation
$$ ((L'_{10} - L'_{31})^2 + (L'_{20} + L'_{23})^2) w(a) = - \kappa^2 w(a),
\tag 7.25 $$
which can be written in the form
$$ \tfrac 1 4 e^{jirs} \hat k_i L'_{rs} e_j{}^{kpq} \hat k_k L'_{pq}
w(a) = - \kappa^2 w(a). \tag 7.26 $$
By means of the procedure used above, for a field with zero
mass and infinite helicity spectrum we find the equation
$$ g_{ik} \Sigma^i \Sigma^k W(x, a) =
g_{ik} \Sigma^{\prime i} \Sigma^{\prime k} W(x, a) =
 \kappa^2 W(x, a).  \tag 7.27 $$

If we consider a field with zero mass and given helicity $M$ based
on the functions (6.11) and (6.12), we see that eqs.\ (7.10) and (7.11)
can be written in the form
$$ \tfrac 1 2 e^{jirs} \hat k_i L'_{rs} w(a) = -i M \hat k^j w(a), 
\tag 7.28 $$
$$ \hat k^i L'_{ik} w(a) = (c - 1) \hat k_k w(a). \tag 7.29 $$
Proceeding as in the other cases, we obtain
$$ \Sigma'_i W(x, a) = - iM A'_i W(x, a), \qquad
   \Sigma_i  W(x, a) = - iM A_i W(x, a),  \tag 7.30 $$
$$ A^i A_{ik} W(x, a) = (c - 1) A_k W(x, a). \tag 7.31 $$

From eqs.\ (3.8), (3.29) and (4.20) we get
$$ (\Omega, \psi(x, a) \psi^{\dagger}(y, b) \Omega) =
\overline W((y, b)^{-1} (x, a )), \tag 7.32 $$
$$ (\Omega, \psi^{\dagger}(y, b) \psi(x, a) \Omega) =
\overline W((-y, b)^{-1} (-x, a )). \tag 7.33 $$
If the function  $\overline W(x, a)$ satisfies a differential equation
invariant under left translations and under the reflection $x \to -x$,
from the Wightman recostruction theorem$^{19, 20}$ we have that
$\psi(x, a)$ satisfies the same equation. In this way we get the field
equations
$$ g^{ik} A_i A_k \psi (x,a) =  {\mu}^2 \psi (x, a), \tag 7.34 $$
$$\tfrac 1 2 g^{ir} g^{ks} A_{ik} A_{rs} \psi (x,a) =
(1 - c^2 - M^2) \psi (x, a), \tag 7.35 $$
$$\tfrac 1 8 e^{ikrs} A_{ik} A_{rs} \psi (x, a) =
-i M c \psi (x, a).  \tag 7.36 $$
For $\mu > 0$, $j > 0$ we have
$$ g_{ik} \Sigma^i \Sigma^k \psi(x, a) = \mu^2 j(j+1) \psi(x, a),
\tag 7.37 $$
for $\mu > 0$, $j = 0$ we have
$$ \Sigma^i \psi (x, a) = 0, \tag 7.38 $$
for $\mu = 0$ and infinite helicity spectrum we have
$$ g_{ik} \Sigma^i \Sigma^k \psi(x, a) = \kappa^2 \psi(x, a),
\tag 7.39 $$
for $\mu = 0$ and helicity $M$ we have
$$ \Sigma_i \psi (x, a) =  iM A_i \psi (x, a),
\tag 7.40 $$
$$ A^i A_{ik} \psi(x, a) = (\overline c - 1) A_k \psi(x, a). \tag 7.41 $$

We have seen in Sec. VI that there is some difficulty in the
definition of fields starting from the function (6.12).  For these fields,
the calculations given above have a purely formal character.

\bigpagebreak
\eject

\subheading {VIII. Flat-Space Theories and Spin-Mass-Shells}

We approximate a small region of the group manifold $\tilde{\Cal P}$
by means of a tangent space with coordinates $x^i$ and $x^{ik} = -x^{ki}$,
which can be identified with the vector space $\Cal T$.
The operators $A_i$ and $A_{ik}$ which appear in the differential equations
satisfied by the fields and by their v. e. v.'s can be replaced by the
partial derivatives with respect to these coordinates.  We get in this way
a field theory in the flat ten-dimensional space and the Fourier transforms
of the fields and of the v. e. v.'s have their support in a Lorentz invariant
manifold defined by  some polynomial equations in the ten-dimensional
``spin-momentum space'' $\Cal T^*$ with coordinates
$k_i$ and $k_{ik} = - k_{ki}$.
These equations are obtained from the field equations by means
of the substitutions  $A_i \to -i k_i, \,\, A_{ik} \to -i k_{ik}$.
To indicate the positive-energy part ($k^0 = - k_0 > 0$) of this
manifold, we use the term ``spin-mass-shell''.  The two-point v. e. v. of
the flat-space theory is the Fourier transform of a positive Lorentz invariant
measure on the spin-mass-shell.  As we shall see, in some cases the
replacement of non-commuting operators by commuting quantities may lead
to inconsistencies or ambiguities. Nevertheless, the correspondence
between theories in $\tilde{\Cal P}$ and in the flat space is an
unavoidable heuristic instrument.
Note that there is no ambiguity in the higher degree terms of the equations
defining the spin-mass-shell. If we drop all the other terms, we get a set of
homogeneous equations that define an unambiguous dilatation invariant
``asymptotic'' manifold.

From eqs.\ (7.7), (7.14) and (7.15) we get the following
equations valid on the spin-mass-shell of all the elementary field theories.
$$ g^{ik} k_i k_k = -\mu^2, \qquad
\tfrac 1 2 g^{ir} g^{ks} k_{ik} k_{rs} = c^2 + M^2 -1, \qquad
\tfrac 1 8 e^{ikrs} k_{ik} k_{rs} = i M c. \tag 8.1  $$
In a similar way from eqs.\ (7.23), (7.24), (7.27), (7.30) and (7.31)
we obtain other equations valid for the various specific cases.

Since the spin-mass-shell is Lorentz-invariant, it is determined by its
intersection with the hyperplane $k_i = \hat k_i$. This intersection can be
considered as a manifold in the six-dimensional ``spin space'' with
coordinates $k_{ik}$ and we call it the ``spin-shell''.  Its dimension
is given by the dimension of the spin-mass-shell minus three.
It is convenient to introduce the three-dimensional vectors
$$ \bold k=(k^1, k^2, k^3),\qquad \bold k'=(k^{23}, k^{31}, k^{12}), \qquad
\bold k''=(k^{10}, k^{20}, k^{30})  \tag 8.2 $$
and similar notations for the coordinates $x^i, x^{rs}$.  Then the last
two conditions of eq.\ (8.1) take the form
$$ (\bold k')^2 - (\bold k'')^2 = c^2 + M^2 - 1,
\qquad   \bold k' \cdot \bold k'' =  i Mc. \tag 8.3 $$

The spin-shell in the case $\mu > 0$ is described by the equations
$$ (\bold k')^2 = j(j+1), \qquad  (\bold k'')^2 = j(j+1) + 1 -c^2 - M^2,
\qquad   \bold k' \cdot \bold k'' =  i Mc. \tag 8.4 $$
Note that the right-hand sides are real and satisfy the inequality
$$ j(j + 1) \left(j(j+1) + 1 -c^2 - M^2 \right) + M^2 c^2 =
(\bold k')^2 (\bold k'')^2 - (\bold k' \cdot \bold k'')^2 \geq 0. \tag 8.5 $$
If $j > 0$ this manifold has dimension three, but if $j = M = 0$ eq.\ (8.4)
takes the simpler form
$$ \bold k' = 0, \qquad  (\bold k'')^2 = 1 -c^2 \tag 8.6 $$
and describes a two-dimensional manifold.
In both cases the spin-shell is compact and the rotation group acts
transitively on it, namely the spin-shell is an orbit of the rotation group
in the spin space. It follows that the spin-mass-shell is an orbit
of the Lorentz group in the spin-momentum space.

The orbits of the rotation or of the Lorentz group which correspond to a
field on the group $\tilde{\Cal P}$ are called ``allowed orbits''.
It is useful to consider all the orbits, not necessarily allowed, which
can be classified by means of the invariants
$(\bold k')^2$, $(\bold k'')^2$ and $\bold k' \cdot \bold k''$
satisfying the condition (8.5). One can also use eq.\ (8.4) to introduce
the parameters $j$, $c$ and $M$ even when they do not
label any i. u. r.  Given an orbit with $\bold k' \cdot \bold k'' \neq 0$,
eq.\ (8.4) determines the real quantities $j$, $M$ and the imaginary
quantity $c$, up to a common change of sign of $M$ and $c$. They satisfy
the conditions
$$ j > 0, \qquad c^2 < 0, \qquad
0 < M^2 \leq j(j+1) \left(1 + (j(j+1) + |c^2|)^{-1}\right). \tag 8.7 $$
Note that only some discrete values of the parameters $j$ and $M$
are allowed and that they are rather uniformly
distributed in the set defined by eq.\ (8.7).
These orbits are three-dimensional in the general case, but have dimension
two when the equality sign holds in eq.\ (8.5) or in the last eq.\ (8.7),
namely when the vectors $\bold k'$ and $\bold k''$ are parallel.
 The two-dimensional orbits correspond to
values of $M$ which are not allowed, but for large $j$
are relatively near to the allowed values $M = \pm j$.

If $\bold k' \cdot \bold k'' = 0$, we have two possible choices of
$M$ and $c$, namely
$$M= 0, \qquad 1 - c^2 =  (\bold k'')^2 - (\bold k')^2, \qquad j \geq 0,
 \tag 8.8 $$
or
$$  c = 0, \qquad  M^2 - 1 = (\bold k')^2 - (\bold k'')^2, \qquad
j(j+1) = (\bold k')^2  \geq M^2 -1. \tag 8.9 $$
There is some ambiguity in the parametrization of the orbits, which disappears
if we consider only allowed orbits and allowed values of the paramaters.
These orbits are three-dimensional in the general case and two-dimensional
if $\bold k' = 0$ or $\bold k'' = 0$.  If both these vectors vanish,
we have a zero-dimensional orbit, corresponding to $j = M = 0$,  $c^2 = 1$,
namely to a Minkowskian theory.

In the case $\mu = 0$, infinite helicity spectrum, the spin-shell is
described by the equations (8.3) and
$$ (k_{10} - k_{31})^2 + (k_{20} + k_{23})^2 = \kappa^2 > 0. \tag 8.10 $$
It has dimension three and it is an unbounded orbit of the little group
$\tilde E(2)$.

In the case $\mu = 0$, helicity $M$, from eq.\ (7.30) we get the equations
$$ k_{10} = k_{31}, \quad k_{20} = - k_{23}, \qquad k_{12} = M. \tag 8.11 $$
Eq. (7.31) contains products of non-commuting operators and gives
ambiguous results. A direct substitution gives
$$ k_{30} = i(1 - c), \tag 8.12 $$
but this result is not compatible with eq.\ (8.3).
We consider the equation
$$ k_{30} = N(c), \tag 8.13 $$
without specifying the function $N(c)$, apart from the conditions $N(1) = 0$
and $N(c) \simeq -ic $ for large imaginary $c$.
If $N(c)$ is real, eqs.\ (8.11) and (8.13) define a two-dimensional
unbounded orbit of $\tilde E(2)$, unless $M = N = 0$.

In order to complete the list of the orbits of $\tilde E(2)$
in the spin space, we have to consider the zero-dimensional trivial orbit,
corresponding to a Minkowskian scalar field,
and a set of one-dimensional bounded orbits defined by the conditions
$$  k_{10} = k_{31}, \quad    k_{20} = - k_{23}, \qquad
k_{12} = k_{30} = 0,  \tag 8.14  \nopagebreak $$
$$ (k_{31})^2 + (k_{23})^2 = \nu^2 > 0.  $$
The corresponding orbits of the Lorentz group are four-dimensional and
have a large symmetry group$^{2, 7}$ locally isomorphic to $L(4, \bold R)$.
It has been shown in ref. 7 that they are not allowed according to
our definition.

Now we study the v. e. v.'s of the flat-space theories with $\mu > 0$.
Given a positive-mass orbit of the Lorentz group in the spin-momentum space
and positive Lorentz-invariant  measure $m$ on it, the v. e. v. can be
written in the form
$$ V(x, \bold x', \bold x'') = (2 \pi)^{-3}
 \int \exp(-i k \cdot x +i \bold k'\cdot \bold x'
-i \bold k''\cdot \bold x'') \, d m = \tag  8.15 \nopagebreak $$
$$ = (2 \pi)^{-3} \int \exp(-i k \cdot x +i \bold k'\cdot \bold x'
-i \bold k''\cdot \bold x'')
\tilde v(\bold{\hat k'}, \bold{\hat k''})\theta(k^0) \delta(k \cdot k + \mu^2)
\, d^4k \, d^3 \bold k' \, d^3 \bold k'', $$
where
$$ \hat k_{ik} = \Lambda^r{}_i(a_k) \Lambda^s{}_k(a_k) k_{rs},  \tag 8.16  $$
$\tilde v(\bold k', \bold k'')\, d^3 \bold k' \, d^3 \bold k''$
represents a positive rotation-invariant measure concentrated on the 
spin-shell and $a_k$ is defined by eq.\ (3.21).

The integral (8.15) can be written in the form
$$ V(x, \bold x', \bold x'') = (2 \pi)^{-3} \int \exp(-i k \cdot x)
v(\bold{\hat x'}, \bold{\hat x''})
\theta(k^0) \delta(k \cdot k + \mu^2) \, d^4k, \tag 8.17 $$
where
$$ \hat x^{ik} = \Lambda_r{}^i(a_k) \Lambda_s{}^k(a_k) x^{rs}  \tag 8.18 $$
and
$$ v(\bold x', \bold x'')  = \int \exp(i \bold k' \cdot \bold x'
-i \bold k'' \cdot \bold x'') \tilde v(\bold k', \bold k'')
\, d^3 \bold k' \, d^3 \bold k''. \tag 8.19 $$

If the vectors $\bold k', \, \bold k''$ represent an arbitrary
point of the spin-shell, we can put
$$ v(\bold x', \bold x'') =
I(\bold k', \bold k'', \bold x', \bold x'')
 = \int_{SO(3)} \exp(i R \bold k' \cdot \bold x'
-i R \bold k'' \cdot \bold x'') \, d^3 R, \tag 8.20 $$
where $R$ is a three-dimensional rotation matrix and $d^3 R$ is the
normalized invariant measure on the rotation group.
In the general case, this integral cannot be expressed in terms of elementary
functions, but if the vectors $\bold k', \, \bold k''$
are parallel or antiparallel we have
$$ v(\bold x', \bold x'')  = t^{-1} \sin t, \tag 8.21 $$
where
$$ t^2 = (\|\bold k'\| \bold x' \mp \|\bold k''\| \bold x'')^2.
 \tag 8.22 $$
If we consider a theory with $j = M = 0$, from eq.\ (8.6) we get
$$ t^2 = (1- c^2) (\bold x'')^2. \tag 8.23 $$

Eq.\  (8.17) is similar to eq.\ (3.24) and it is interesting to compare
the functions $v(\bold x', \bold x'')$ and $w(a)$, where $a$ is given by the
exponential
$$a = \exp(\tfrac 1 2 (\bold x'' -i \bold x') \cdot \bold\sigma). \tag 8.24 $$
Eq.\ (A.27) gives the power expansion of the integral (8.20) up to quadratic
terms in the variables $\bold x', \bold x''$. A comparison with eq.\ (A.26)
shows that, disregarding higher order terms, we have
$$ w(a) \simeq v(\bold x', \bold x''). \tag 8.25  $$
It follows that if $f(x)$ is a test function and $a$
is given by eq.\ (8.24), we have
$$ \int f(x) W(x, a) \, d^4 x  \simeq
\int f(x) V(x, \bold x', \bold x'') \, d^4 x  \tag 8.26 $$
up to terms of the second order.

In order to simplify the integral (8.17), we proceed as in Sec. IV,
namely we put
$$ \bold x' = (0, 0, \alpha), \qquad \bold x'' = (0, 0, \beta) \tag 8.27 $$
and we use the definitions (4.23), (4.24) and (4.25). From eq.\ (8.18)
we obtain
$$ \hat{\bold x}' = (0, -\beta \sinh\eta, \alpha \cosh\eta)
= \mu^{-1}(0, -\beta p, \alpha q), \tag 8.28 \nopagebreak $$
$$ \hat{\bold x}'' = (0, \alpha \sinh\eta, \beta \cosh\eta) =
\mu^{-1}(0, \alpha p, \beta q). $$
The analogue of eq.\ (4.28) is
$$ V(x, \alpha, \beta) = (2 \pi)^{-1} \int_0^{\infty}
\Delta(q \sigma, \epsilon(x^0))
 J_0(p \rho) v(\bold x', \bold x'') p \, d p \tag 8.29 $$
and in a similar way one writes the analogues of eqs.\  (4.31) and (4.32)
which determine the commutator.

If in eq.\ (8.20) we replace the integrand by the maximum of its modulus,
we get the inequality
$$ |I(\bold k', \bold k'', \bold x' +i\bold y', \bold x'' +i\bold y'')| \leq
\tag 8.30  \nopagebreak $$
$$ \leq \exp\left((\bold k')^2 (\bold y')^2 + (\bold k'')^2(\bold y'')^2
+2 (\bold k' \cdot \bold k'') (\bold y' \cdot \bold y'')
+2 \| \bold k' \times \bold k''\| \|(\bold y' \times \bold y''\|
\right)^{\tfrac 1 2}.  $$
For large $|p|$ we have $\text{Im}\,q \simeq \text{Im}\,p$ and, with this
approximation, from eq.\ (8.28) we obtain
$$ |v(\hat{\bold x}', \hat{\bold x}'')| \leq
\exp\left(\mu^{-1} |\text{Im}\,p| (\alpha^2 + \beta^2)^{\tfrac 1 2}
((\bold k')^2 + (\bold k'')^2 +2 \| \bold k' \times \bold k'' \|)^{\tfrac 1 2}
\right). \tag 8.31  $$
If we take into eccount eq.\ (4.33) we see that in the analogue of eq.\ (4.32)
we can close the integration path at infinity in the upper half plane if
$$ \rho > \sigma + \mu^{-1} (\alpha^2 + \beta^2)^{\tfrac 1 2}
((\bold k')^2 + (\bold k'')^2 +2 \| \bold k' \times \bold k'' \|)^{\tfrac 1 2}
\tag 8.32 $$
and under this condition the commutator of the flat-space theory vanishes.
We remark that the flat-space theory has stronger local commutation properties
than the corresponding theory on $\tilde{\Cal P}$.

For $j = M = 0$, we can use eqs.\ (8.21) and (8.23) and from eq. (8.28)
we have
$$ t^2 = (1- c^2) (\alpha^2 (\sinh\eta)^2 + \beta^2 (\cosh\eta)^2) =
(1 - c^2) (\beta^2 + (\alpha^2 + \beta^2) \mu^{-2} p^2). \tag 8.33 $$
If we remark that, for small values of $\alpha$ and $\beta$, eq.\ (5.25)
gives
$$ \zeta^2 \simeq \alpha^2 (\sinh\eta)^2 + \beta^2 (\cosh\eta)^2, \tag8.34 $$
we can easily verify eq.\ (8.26) in this special case.

\bigpagebreak

\subheading {IX. A Theory with Broken $Sp(4, \bold R)$ Symmetry}

Now we consider a scalar flat-space theory invariant with respect to a group
$\Cal F$ larger than $SL(2, \bold C)$. It is defined  by a spin-mass-shell
which is an orbit of $\Cal F$ and can be decomposed into orbits of the
Lorentz group. As a consequence, its two-point v. e. v. is
a superposition (an integral) of the Lorentz invariant v. e. v.'s
$V(x, \bold x', \bold x'')$ described in the preceding Section.
If all the Lorentz orbits which appear in the decomposition are allowed
(apart from a set of vanishing measure),
we can consider the analogous superposition of the v. e. v.'s $W(x, a)$
and we find the v. e. v. of a non-elementary theory on $\tilde {\Cal P}$
which corresponds, in some sense, to the $\Cal F$-invariant theory
on the flat space. In general the theory on $\tilde {\Cal P}$
has lost the symmetry under the large group $\Cal F$, but some
consequence of this higher symmetry remains in the short-distance limit.

In refs.\ 7, 30 we have described several flat-space theories symmetric
with respect to $SL(4, \bold R)$ or to one of its subgroups isomorphic to
$Sp(4, \bold R)$, which form a one-parameter family$^2$. The positivity of the
energy requires that the spin-mass-shell is contained in a closed invariant
cone ${\Cal T}^{*+}$, the dual of the cone ${\Cal T}^+$ that describes the
causal properties of the theory. Therefore, only the allowed Lorentz orbits
contained in ${\Cal T}^{*+}$ are interesting for the purpose we are 
discussing. The corresponding spin-shells must be bounded and 
this requirement
excludes all the zero-mass allowed theories considered in the preceding
Section, but the Minkowskian zero-mass theory coresponding to the spin-shell
reduced to the origin. In addition to this theory,
only positive-mass elementary theories can be used in the
construction of a theory on $\tilde {\Cal P}$ with broken higher symmetry.
If we look at the definition of the cone ${\Cal T}^{*+}$ $^7$, we see that
for $\mu > 0$ the spin-shell must be contained in the set defined by
$$ (\bold k')^2 + (\bold k'')^2  + 2 \| \bold k' \times \bold k'' \|
\leq \mu^2 \tag 9.1 $$
(see the Appendix B).
This inequality gives rise to a complicated constraint on the parameters
$j$, $M$, and $c$, in particular we get
$$ j(j + 1) + 1 - c^2 \leq \mu^2. \tag 9.2 $$
The mass $\mu$ is measured in natural units of the order of the Planck mass
and for the observable particles it is very small. It follows that we must
have $j = M = 0$ and $1 - c^2 \ll 1$, namely the theory must be very near to
the Minkowskian limit. We also see that the particles with small $\mu$ and
non-vanishing spin cannot be described by scalar fields.  We have already
remarked in Section 5 that spin has two different origins: in this case the
spin is generated by the field indices, as in eqs.\ (5.15) and (6.28).
Combined with eq.\ (8.32),
eq.\ (9.1) ensures that the commutator of the flat-space theory
vanishes for
$$ \rho > \sigma + (\alpha^2 + \beta^2)^{\tfrac 1 2}. \tag 9.3 $$

We consider the spin-mass-shells described in refs.\ 7, 30 and we exclude
the four-dimensional one, composed of a single zero-mass not allowed
Lorentz orbit. The decomposition of these spin-mass-shells, (disregarding
a set of vanishing measure) contains only positive-mass Lorentz orbits.
In general not all these orbits are allowed, but one can try to replace
the integral over the continuous parameters $j$ and $M$ by a sum over the
discrete allowed values. In this way, we may also obtain a discrete mass
spectrum.  This procedure will be examined elsewhere; in
the following we consider a particular choice of the group $\Cal F$ isomorphic
to $Sp(4, \bold R)$ and a particular class of orbits, in such a way that
all the Lorentz orbits that appear in the decomposition are allowed.

For the description of the spin-mass-shells invariant with respect to
$Sp(4, \bold R)$, locally isomorphic to the anti-de Sitter group
$SO^{\uparrow}(2,3)$, it is convenient to introduce the notation
$$ k^{5i} = - k^{i5} = k^i, \qquad i = 0, 1, 2, 3 \tag 9.4 $$
and to consider the quantities $k^{uv}$ as the components of an antisymmetric
tensor in a five-dimensional space  with metric $g_{55} = g_{00} = -1, \quad
g_{11} = g_{22} = g_{33} = 1$.  In the following the indices 
$u, v, w, x, y, z$ take the values $5, 0, 1, 2, 3$.

We consider the orbit which contains the point defined by $k^0 = s \geq 0$,
$\bold k = \bold k' = \bold k'' = 0$. Then it also contains the Lorentz orbit
that corresponds to a Minkowskian theory with mass $\mu = s$. It is
six-dimensional and in ref.\ 30 it has been called
$\Cal O_{4, \tfrac 1 2 s, \tfrac 1 2 s}$ for $ s > 0$,
or $\Cal O_{2,0}$ for $s = 0$.  The following $O(2, 3)$-invariant set of
conditions is satisfied on the orbit for all the values of $s$:
$$ e^{uvwxy} k_{vw} k_{xy} = 0.   \tag 9.5 $$
In the four-dimensional formalism these conditions take the form
$$ e^{ikrs} k_{ik} k_{rs} = 0,   \qquad
e^{ikrs} k_k k_{rs} = 0 \tag 9.6 $$
and in the three-dimensional formalism we can write
$$ \bold k \cdot \bold k' = 0, \qquad \bold k' \cdot \bold k'' = 0, 
\tag 9.7 $$
$$  \bold k'  = (k^0)^{-1} \bold k'' \times \bold k.  \tag 9.8 $$
It is clear that eq\. (9.8) implies eq\. (9.7) and therefore the
whole set of conditions (9.5). We see that these conditions are not
independent and define a seven-dimensional manifold.

The $O(2, 3)$-invariant manifold defined by eq.\ (9.5) can be parametrized
by means of the coordinates $\bold k, \bold k'', k^0$ and it is easy to
control that the measure defined by
$$  (k^0)^{-2} d^3 \bold k \, d^3 \bold k'' \, dk^0 \tag 9.9 $$
is invariant under $O(2, 3)$.    In order to get an orbit of
$SO^{\uparrow}(2, 3)$, we have to introduce the further invariant
condition
$$ \tfrac 1 2 g^{ux} g^{vy} k_{uv} k_{xy} =
(k^0)^2 -(\bold k)^2 + (\bold k')^2 - (\bold k'')^2 = s^2, \tag 9.10 $$
that, together with eq\. (9.8) gives
$$ k^0 = \tag 9.11 \nopagebreak $$
$$ = \pm \left(\tfrac 1 2 ((\bold k)^2 + (\bold k'')^2 + s^2)
\pm\tfrac 1 2  \left(((\bold k)^2 + (\bold k'')^2 + s^2)^2
- 4 \| \bold k'' \times \bold k \|^2
   \right)^{\tfrac 1 2} \right)^{\tfrac 1 2}. $$
This formula describes the orbit we are considering if we choose
the sign $+$ twice.

From eqs. (9.8) and (9.10) we see that in the decomposition of this orbit
into orbits of the Lorentz group, we find,
besides the above mentioned orbit with mass $\mu = s$, the
allowed orbits labelled, in accord with eq.\ (8.4), by the parameters
$$ M = j = 0, \quad \mu^2 = 1 - c^2 + s^2 > s^2. \tag 9.12  $$

In order to find an invariant measure on this orbit, we have to multiply
eq\. (9.9) by the appropriate invariant $\delta$-function
and integrate over $d k^0$. The result is
$$  d m_s = \delta\left((k^0)^2 -(\bold k)^2 + (\bold k')^2
- (\bold k'')^2 - s^2 \right) (k^0)^{-2} d^3 \bold k \, d^3 \bold k'' 
\, dk^0 = \tag 9.13 \nopagebreak $$
$$ = \tfrac 1 2 (k^0)^{-1} \left((k^0)^2 -(\bold k')^2\right)^{-1} \,
d^3 \bold k \, d^3 \bold k'', $$
where $\bold k'$ and $k^0$ are given by eqs\. (9.8) and (9.11).

The Fourier transform of this measure can be performed in two steps:
$$ V_s(x, \bold x', \bold x'') = (2 \pi)^{-3} \int \exp(-i k \cdot x)
A(k^0, \bold k, \bold x', \bold x'') \, d^4 k, \tag 9.14 $$
$$ A(k^0, \bold k, \bold x', \bold x'') =  (k^0)^{-2} \int
\exp(i \bold k' \cdot \bold x' -i \bold k'' \cdot \bold x'') \times
\tag 9.15 \nopagebreak $$
$$\times \theta\left(k^0 - 2^{-\tfrac 1 2} ((\bold k)^2 +
(\bold k'')^2 + s^2)^{\tfrac 1 2}\right)
\delta\left((k^0)^2 - (\bold k)^2 + (\bold k')^2 - (\bold k'')^2 - s^2\right)
\, d^3 \bold k''. $$
If $\bold k = 0$, we also have $\bold k' = 0$ and therefore,
$$ A(k^0, 0, \bold x', \bold x'') = \tag 9.16 \nopagebreak $$
$$= \theta(k^0 - s) (k^0)^{-2} \int \exp(-i \bold k'' \cdot \bold x'')
\delta((\bold k'')^2 - (k^0)^2 + s^2) \, d^3 \bold k'' =
\nopagebreak $$
$$ = 2 \pi (k^0)^{-2} ((k^0)^2 - s^2)^{\tfrac 1 2} \theta (k^0 - s)
v^{0c0}(\bold x''), $$
where $v^{0c0}(\bold x'')$ is given by eqs.\ (8.21) and (8.23) with
$1 - c^2 = \mu^2 - s^2$.  In general, by means of the Lorentz transformation
 $a_k$ we obtain
$$ A(k^0, \bold k, \bold x', \bold x'') =
2 \pi \mu^{-2} (\mu^2 - s^2)^{\tfrac 1 2} \theta (\mu^2 - s^2) \theta(k^0)
v^{0c0}(\hat{\bold x}''), \tag 9.17 $$
where $\hat{\bold x}''$ is given by eq. (8.18) and
$\mu^2 = (k^0)^2 - (\bold k)^2$.
In conclusion, we have
$$ V_s(x, \bold x', \bold x'') = \tag 9.18 \nopagebreak $$
$$ = (2 \pi)^{-2} \int \exp(-i k \cdot x) \mu^{-2} (\mu^2 - s^2)^{\tfrac 1 2}
\theta (\mu^2 - s^2) \theta(k^0) v^{0c0}(\hat{\bold x}'') \, d^4 k.  $$

In order to obtain the v. e. v.'s of the theory defined
on $\tilde{\Cal P}$, we have just to replace $v^{0c0}$ by $w^{0c0}$
and we get
$$ W_s(x, a) = \tag 9.19 \nopagebreak $$
$$ = (2 \pi)^{-2} \int \exp(-i k \cdot x) \mu^{-2} (\mu^2 - s^2)^{\tfrac 1 2}
\theta (\mu^2 - s^2) \theta(k^0) w^{0c0}(a_k^{-1} a a_k) \, d^4 k.  $$

Also in this case we can introduce the parameters $\alpha$ and $\beta$,
parametrize $k$ as in eq.\ (4.23) and perform the integration over the
variables $\xi$ and $\psi$. We obtain
$$ V_s(x, \alpha, \beta)= 2 \int_0^{\infty} \int_0^{\infty}
(q^2 - p^2)^{-1} (q^2 - p^2 - s^2)^{\tfrac 1 2} \times \tag 9.20 
\nopagebreak $$
$$ \times \theta(q^2 - p^2 - s^2) \Delta(q \sigma, \epsilon(x^0))
 J_0(p \rho) t^{-1} \sin t \, p dp \, q dq,  $$
where
$$t^2 = (q^2 - p^2)^{-1} (q^2 - p^2 -s^2)(p^2 \alpha^2 + q^2 \beta^2)
\tag 9.21 $$
and
$$ W_s(x, u_3(\alpha) a_3(\beta)) = 2 \int_0^{\infty}\int_0^{\infty}
(q^2 - p^2)^{-1} (q^2 - p^2 - s^2)^{\tfrac 1 2} \times \tag 9.22 
\nopagebreak $$
$$ \times \theta(q^2 - p^2 - s^2) \Delta(q \sigma, \epsilon(x^0))
 J_0(p \rho) (c \sinh\zeta)^{-1} \sinh(c \zeta) \,p dp \,q dq, $$
where
$$ c^2 = s^2 - q^2 + p^2 +1, \qquad
\cosh\zeta = (q^2 - p^2)^{-1} (q^2 \cosh\beta -p^2 \cos\alpha). \tag 9.23 $$

The field defined by $W_s$ satisfies the equations
$$ e^{uvwxy} A_{vw} A_{xy} \psi = 0,   \tag 9.24 $$
$$ \tfrac 1 2 g^{ux} g^{vy} A_{uv} A_{xy} \psi = - s^2 \psi, \tag 9.25 $$
written in the five-dimensional formalism. In fact, it is a direct integral
of fields that satisfy eqs.\ (7.34)-(7.36) and (7.38)
with the parameters constrained
by eq.\ (9.12). This field describes ``particles'' with vanishing spin and
a continuous  mass spectrum lying on the half-line $\mu \geq s$.

The function $V_s$ has been computed in ref.\ 30 and, with the conventions
adopted here, is given by
$$ V_s(x, \bold x', \bold x'') = \lambda_1^{-1} \lambda_2^{-1}
(\lambda_1 + \lambda_2)^{-1} \exp(-\tfrac 1 2 s (\lambda_1 + \lambda_2)),
\tag 9.26 $$
where $\lambda_1, \lambda_2$ are given by
$$ \lambda_{1,2}^2 = A \pm (A^2 - B)^{\tfrac 1 2}, \tag 9.27 $$
$$ A = - \sigma^2 + \rho^2 - \alpha^2 + \beta^2, \tag 9.28 $$
$$ B = (\sigma^2 - \rho^2 - \alpha^2 - \beta^2)^2 -
   4 \rho^2 (\alpha^2 + \beta^2)  $$
(see the Appendix B for more details).
The signs of  $\lambda_1, \lambda_2$ are determined in such a way that
their real parts are positive or, if one of them vanishes,
it becomes positive after the
addition of a small positive imaginary part to $x^0$.

Starting from eqs.\ (9.20)-(9.23), and introducing the new integration
variables $p' = \epsilon p, \,\, q' = \epsilon q$, it is easy to prove that
$$ \lim_{\epsilon \to 0} \left(\epsilon^3 W_s(\epsilon x, u_3(\epsilon\alpha)
a_3(\epsilon\beta))\right) = V_0(x, \alpha, \beta),  \tag 9.29 $$
$$ \lim_{\epsilon \to 0} \left(\epsilon^3 V_s(\epsilon x, \epsilon\alpha,
\epsilon\beta)\right) = V_0(x, \alpha, \beta).  \tag 9.30 $$
This means that in the short-distance limit, namely near to the unit of
the group, the v. e. v. $W_s$ coincides with the distribution
$V_0$, symmetric with respect to $Sp(4, \bold R)$. In other words, the
symmetry broken by the structure of the Poincar\'e group survives in
the short distance limit. We also see that in the short-distance limit
the dependence on the parameter $s$ disappears.

\bigpagebreak

\subheading {X. Radiation from an Accelerated Source}

The simplest exercise with a free quantum field  is its interaction
with an external source. We consider a scalar Hermitean field
and we write the time integral of the interaction Hamiltonian in the form
$$ F = \int H'(t) \, dt = \int f(x, a) \psi (x, a) \, d^4 x \, d^6 a.
\tag 10.1 $$
The scattering operator is given by
$$ S = \exp(-i F) =
 \exp(- \tfrac 1 2 \| F \Omega\|^2)  :\exp(-i F): \tag 10.2 $$
(time ordering is not necessary, since it introduces only an over all
 phase factor). The number of emitted particles follows a Poisson
distribution with average  value
$$ \langle n \rangle =  \| F \Omega\|^2.  \tag 10.3 $$
This is just the quantity given by eq.\ (3.12).

In general, the function $f$ that describes the source is not arbitrary;
it may be subject to some conservation law or to some other constraint
arising from the field equations. For instance, it is not clear if a
point particle has to be be described by a one-dimensional trajectory
in $\tilde{\Cal P}$ or by
a manifold with higher dimension as it is discussed in ref. 31.
 The formulation of possible constraints
requires a deeper understanding of the theory and we disregard this
problem in the following exercise. We consider a source that is
bound to an accelerated frame, obtained from an initial frame by means
of the following one-parameter group of Poincar\'e transformations:
$$ t \to \left((a^{-1} \sinh(at), 0, 0, a^{-1} (\cosh(at) -1)), 
a_3(at)\right). \tag 10.4 $$
An infinitesimal transformation of this group is the product of a time
translation by an infinitesimal amount $dt$ and a boost along the $z$
axis with infinitesimal  velocity $a dt$,  
where $a$ represents a constant acceleration.
The parameter $t$ is the proper time of the accelerated frame.

Since a point source is too sigular, we consider
a source concentrated on a disk lying in the $x^1, x^2$ plane
of the accelerated frame; then we have:
$$ F = \int P(r) Q(t) \times \tag 10.5 \nopagebreak $$
$$ \times \psi\left((a^{-1} \sinh(at), r \cos\phi, r \sin\phi,
a^{-1} (\cosh(at) -1)), a_3(at)\right) \, r dr \, d\phi \, dt, $$
where $P(r)$ is the ``density'' of the disk and $Q(t)$ is a function equal to
one in an interval of lenght $T$ and going smoothly to zero outside this
interval. We expect that when $T$ is large, $\langle n \rangle$ is 
proportional to $T$.

Then from eq.\ (10.3) we obtain
$$ \langle n \rangle = \tag 10.6 \nopagebreak $$
$$ = \int P(r_1) P(r_2) Q(t_1) Q(t_2)
W(x, a_3(at_2 - at_1)) \, r_1 dr_1 \, d\phi_1 \, r_2 dr_2 \, d\phi_2
\, dt_1 \, dt_2, $$
where
$$ x^0 = a^{-1} \sinh(at), \qquad  x^3 = a^{-1} (\cosh(at) - 1), \qquad
t = t_2 - t_1,    \tag 10.7 \nopagebreak $$
$$ x^1 = r_2 \cos \phi_2 - r_1 \cos \phi_1,     \qquad
   x^2 = r_2 \sin \phi_2 - r_1 \sin \phi_1.  $$
Note that $W$ depends on the quantities (4.5), that take the form
$$ \rho^2 = r_1^2 + r_2^2 -2 r_1 r_2 \cos(\phi_2 - \phi_1), \tag 10.8 $$
$$  \sigma = 2 a^{-1} \sinh(\tfrac 1 2 at). \tag 10.9 $$

We can also write
$$ \langle n \rangle = 2 \pi \int \hat P(\rho) \hat Q(t)
W(x, a_3(\beta)) \, \rho d \rho \, dt, \tag 10.10 $$
where
$$ x = (\sigma, \rho, 0, 0), \qquad \beta = at,  \tag 10.11 $$
$$ \hat P(\rho) = \int P(r)
P(\rho^2 + r^2 +2 \rho r \cos\phi) \, r dr \, d \phi, \tag 10.12 $$
$$ \hat Q(t) = \int Q(t_1) Q(t + t_1) \, dt_1. \tag 10.13 $$
When the proper time interval $T$ is large, we have
$$  T^{-1} \hat Q(t) \simeq 1 - |t| T^{-1}. \tag 10.14 $$
It follows that in the same limit the average number of produced particles
per unit of proper time is given by
$$ T^{-1} \langle n \rangle = 2 \pi \int_{-\infty}^{\infty}\,dt
\int_0^{\infty} \hat P(\rho)
W(x, a_3(\beta)) \, \rho d \rho, \tag 10.15 $$
if the integral converges.  If we take eq.\ (3.29) into account, we can
write
$$ T^{-1} \langle n \rangle = 4 \pi \int_0^{\infty}\,dt
\int_0^{\infty} \hat P(\rho)
\text{Re}\, W(x, a_3(\beta)) \, \rho d \rho. \tag 10.16 $$

Now we consider the theory defined in Sec. IX and we look for singularities
of the integral (10.16). Since the singularities arise for small values of
$t$ and $\rho$, we approximate the function $W$ by means of eqs.\
(9.26)-(9.29), namely we use the formula
$$ W(x, a_3(\beta)) \simeq \lambda_1^{-1} \lambda_2^{-1}
(\lambda_1 + \lambda_2)^{-1} = (\lambda_1^2 - \lambda_2^2)^{-1}
(\lambda_2^{-1} - \lambda_1^{-1}) =  \tag 10.17 \nopagebreak $$
$$ = (4 \rho \beta)^{-1} \left(((\rho - \beta)^2 - \sigma^2)^{-\tfrac 1 2}
- ((\rho + \beta)^2 - \sigma^2)^{-\tfrac 1 2} \right), $$
where $\beta$ and $\sigma$ are given as functions of $t$ by eqs.\
(10.9) and (10.11).

After some calculation, we obtain
$$ \int \hat P(\rho) \text{Re}\, W(x, a_3(\beta)) \, \rho d \rho =
\tag 10.18 \nopagebreak $$
$$ = (4 \beta)^{-1} \int_{\sigma}^{\infty}
(\hat P(r+\beta) - \hat P(|r-\beta|)) (r^2 - \sigma^2)^{-\tfrac 1 2} \, dr +
\nopagebreak $$
$$ + \theta(\beta - \sigma) (2 \beta)^{-1} \int_{\sigma}^{\beta}
\hat P(\beta-r) (r^2 - \sigma^2)^{-\tfrac 1 2} \, dr. $$
If $\hat P(\rho)$ has a bounded derivative, the first integral in the
right hand side has at most a logarithmic singularity for small $t$.
The second integral is present only if $a > 1$ and for small $t$ it behaves as
$$ \hat P(0) (2at)^{-1} \log\left(a + (a^2-1)^{\tfrac 1 2}\right).
\tag 10.19 $$
As a consequence, if $a > 1$, the integral over $t$ in eq.\ (10.16)
diverges.  In other words, if the
acceleration $a$ is larger than a critical value, conventionally taken
equal to one, the number of particles and the energy radiated per unit
proper time become infinite.

\bigpagebreak

\subheading {Appendix A: Properties of the functions $w^{Mcj}(a)$}

The quantities defined by eqs.\ (5.14) or (5.18) are elementary functions,
but for large values of $j$ they are too complicated and it is
preferable to introduce some integral representations. We start from the
realization$^{22-25}$ of the representation $D^{M c}$ by means of operators
acting on the square integrable functions defined on $SU(2)$ which have the
covariance property
$$ f(u_3(\phi) u) = \exp(-i M \phi) f(u). \tag A.1 $$
An orthonormal basis in this Hilbert space is given by the functions
$$ f_{jm}(u) = (2j + 1)^{\tfrac 1 2} R^j_{Mm}(u). \tag A.2 $$
The invariant measure $d^3 u$ on $SU(2)$ is normalized in such a way that
the measure of the whole group is one.

We consider the decomposition
$$ a = k(a) a_0,  \tag A.3  $$
where
$$ a = \left( \matrix a_{11} & a_{12} \\ a_{21} & a_{22} \endmatrix \right)
\in SL(2, \bold C), \tag A.4 $$
$$ k(a) = \left( \matrix (p(a))^{-1} & q(a) \\ 0 & p(a) \endmatrix \right),
\qquad p(a) > 0,  \tag A.5 $$
$$ a_0 = \left( \matrix \alpha & \beta \\ -\overline\beta & \overline\alpha
\endmatrix \right) \in SU(2), \qquad |\alpha|^2 + |\beta|^2 = 1.  \tag A.6 $$
We see that
$$ p(a) = (|a_{21}|^2 + |a_{22}|^2)^{\tfrac 1 2}, \tag A.7 $$
$$ \alpha = (p(a))^{-1} \overline a_{22}, \qquad
\beta = - (p(a))^{-1} \overline a_{21}. \tag A.8 $$
We shall use the properties
$$ p(au) = p(a), \qquad (au)_0 = a_0 u, \qquad  u \in SU(2), \tag A.9 $$
$$ p(u_3(\phi)a) = p(a), \qquad   (u_3(\phi)a)_0 = u_3(\phi) a_0. \tag A.10 $$

The representation operator is defined by
$$ [D^{M c}(a) f](u) = (p(ua))^{2c - 2} f((ua)_0) \tag A.11 $$
and the matrix elements we need are given by
$$ D^{M c}_{jmjm}(a) = (f_{jm}, D^{M c}(a) f_{jm}) =
\tag A.12 \nopagebreak $$
$$ = (2j + 1) \int_{SU(2)} \overline R^j_{Mm}(u)
(p(ua))^{2c -2} R^j_{Mm}((ua)_0) \, d^3 u. $$
If we sum over $m$ and we use eq\. (A.9), we get the required formula
$$ w^{M c j}(a) = \int_{SU(2)} (p(ua))^{2c -2}
R^j_{MM}((uau^{-1})_0) \, d^3 u. \tag A.13 $$

We can also use the integral representation$^{26}$
$$ R^j_{MM}(u) = (2 \pi)^{-1} \int_0^{2 \pi}
F^{Mj}(u_3(-\phi) u u_3(\phi)) \, d \phi, \tag A.14 $$
where
$$ F^{Mj}(u) = (\alpha - \overline \beta)^{j+M}
(\overline \alpha + \beta)^{j-M}    \tag A.15  $$
and we have used the expression (A.6) for the matrix $u$. If we substitute
eq\. (A.14) into eq\. (A.13) and we use the properties (A.9) and (A.10),
we obtain
$$ w^{M c j}(a) = \int_{SU(2)} (p(ua))^{2c -2}
F^{Mj}((uau^{-1})_0) \, d^3 u \tag A.16 $$
or, more explicitly,
$$ w^{M c j}(a) =  \tag A.17 \nopagebreak  $$
$$ = \int_{SU(2)} (|b_{21}|^2 + |b_{22}|^2)^{c -1-j}
(\overline b_{22} + b_{21})^{j + M}
(b_{22} - \overline b_{21})^{j - M} d^3 u,  $$
where
$$ b = u a u^{-1}.  \tag A.18 $$

The exponential mapping can be written in the form
$$ a = \exp(\tfrac 1 2 (\bold x'' -i \bold x') \cdot \bold\sigma) =
\cosh \chi + \tfrac 1 2 \chi^{-1} \sinh \chi \,\,
(\bold x'' -i \bold x') \cdot \bold\sigma, \tag A.19 $$
where
$$ \chi^2 =  \tfrac 1 4 (\bold x'' -i \bold x') \cdot
(\bold x'' -i \bold x'). \tag A.20 $$
If we indicate by $R(u)$ the $SO(3)$ rotation matrix corresponding to
the element $u \in SU(2)$ and we put
$$ \bold y' = R(u) \bold x', \qquad  \bold y'' = R(u) \bold x'',
\tag A.21 $$
we have
$$ b = u a u^{-1} = \cosh \chi + \tfrac 1 2 \chi^{-1} \sinh \chi \,\,
(\bold y'' -i \bold y') \cdot \bold\sigma. \tag A.22 $$
In particular, it is
$$ b_{21} = \tfrac 1 2 ( y''_1 -i  y'_1 +i  y''_2 +  y'_2)
\chi^{-1} \sinh \chi ,    \tag A.23  \nopagebreak $$
$$ b_{22} = \cosh \chi + \tfrac 1 2 (-y''_3 +i  y'_3)
\chi^{-1} \sinh \chi.  $$

If we substitute these expressions into eq\. (A.17) and we disregard terms
of order higher that the second in  the variables $\bold x'$ and 
$\bold x''$, we can perform the integral by means of the formulae
$$ \int_{SU(2)} \bold y' \, d^3 u = 0, \tag A.24 $$
$$ \int_{SU(2)}  y'_r  y''_s \, d^3 u = \tfrac 1 3  \delta_{rs} \,
\bold x' \cdot \bold x''  \tag A.25 $$
and other similar consequences of eq\. (A.21). The result is
$$ w^{M c j}(a) \simeq  \tag A.26  \nopagebreak $$
$$ \simeq 1 - \tfrac 1 6 j(j+1) (\bold x')^2
- \tfrac 1 6 (j(j+1) + 1 - c^2 - M^2) (\bold x'')^2
+ \tfrac 1 3 i M c \,\, \bold x' \cdot \bold x''. $$

Now we want to compare the integral (A.17) with the integral (8.20)
We can expand the exponential in eq.\ (8.20) keeping terms up to the
second order in the variables $\bold x'$ and $\bold x''$ and perform
the integral by means of eqs\. (A.24) and (A.25). We obtain
$$ I(\bold k', \bold k'', \bold x', \bold x'') \simeq
1 - \tfrac 1 6 (\bold k')^2 (\bold x')^2
- \tfrac 1 6 (\bold k'')^2 (\bold x'')^2 +
\tfrac 1 3 \bold k' \cdot \bold k'' \, \bold x' \cdot \bold x'' \tag A.27 $$
and eq.\ (8.25) follows immediately.

Another interesting limit can be derived from eqs.\ (A.17) and (A.23):
$$ \lim_{n \to \infty} w^{nM, nc, nj}
\left(\exp((2n)^{-1}(\bold x'' -i \bold x') \cdot \bold\sigma)\right) =
\tag A.28 \nopagebreak $$
$$ = \int_{SU(2)} \exp((j-c) y_3'')
\exp(\tfrac 1 2 (j+M)(-y''_3 -i  y'_3 + y''_1 -i  y'_1 +i  y''_2 + y'_2))
\times \nopagebreak $$
$$ \times \exp(\tfrac 1 2 (j-M)(-y''_3 +i  y'_3 - y''_1 -i  y'_1 +i  y''_2
- y'_2)) \, d^3 u = I(\bold k', \bold k'', \bold x', \bold x''), $$
where
$$ \bold k' = (-j, -iM, -M), \qquad
  \bold k'' = (iM, -j, -ic). \tag A.29 $$
The quantity $I(\bold k', \bold k'', \bold x', \bold x'')$ defined in eq.\
(8.20) is an entire
analytic function of its arguments and it depends on $\bold k',\, \bold k''$
through the invariants
$$ (\bold k')^2 = j^2, \qquad   (\bold k'')^2 = j^2 -c^2 -M^2 , \qquad
   \bold k' \cdot \bold k'' = iMc. \tag A.30 $$

If we indicate by $\bold k'_n, \bold k''_n$ the coordinates of a
representative point of the orbit defined by the parameters $nM, nc, nj$,
we have
$$ \lim_{n \to \infty} v^{nM, nc, nj}(n^{-1} \bold x', n^{-1} \bold x'') =
\lim_{n \to \infty}
I(\bold k'_n, \bold k''_n, n^{-1} \bold x', n^{-1} \bold x') =
\tag A.31 \nopagebreak $$
$$ = \lim_{n \to \infty}
I(n^{-1} \bold k'_n, n^{-1} \bold k''_n, \bold x', \bold x') =
I(\bold k', \bold k'', \bold x', \bold x'), $$
where
$$ \bold k' = \lim_{n \to \infty}(n^{-1} k'_n), \qquad
   \bold k'' = \lim_{n \to \infty}(n^{-1} k''_n).  \tag A.32 $$
Since these limits satisfy eq.\ (A.30), we see that the limits (A.28) and
(A.31) are equal.  This result can be used to generalize the treatment of
Sec. IX to a larger class of theories with a broken higher symmetry.

\bigpagebreak

\subheading {Appendix B: Geometry of the vector spaces
$\Cal T$ and $\Cal T^*$}

In this Appendix we summarize some results of refs.\ 2, 3, 7.
We use the Dirac matrices with the properties
$$ \gamma_i \gamma_k + \gamma_k \gamma_i = 2 g_{ik}, \qquad
 \gamma_k{}^T = -C^{-1} \gamma_k C, \qquad C^T = -C. \tag B.1 $$
We adopt the Majorana representation in which the Dirac matrices
are real and we put $C = \gamma_0$.
The vectors of $\Cal T$ and $\Cal T^*$ can be labelled, respectively,
by means of the real symmetric $4 \times 4$ matrices
$$ \hat x= \tfrac 1 2 x^k C^{-1} \gamma_k -
\tfrac 1 4 x^{rs} C^{-1} \gamma_r \gamma_s,
 \tag B.2  $$
$$ \hat k = - \tfrac 1 2 k_k \gamma^k C +
\tfrac 1 4 k_{rs} \gamma^r \gamma^s C.
 \tag B.3  $$
The closed cones $\Cal T^+$ and $\Cal T^{*+}$ contain the elements labelled
by positive semidefinite matrices.

The following formulas are useful:
$$ \text{Tr}\,(\hat k \hat x) = -k_k x^k + \tfrac 12 k_{rs} x^{rs} =
k^0 x^0 - \bold k \cdot \bold x + \bold k'\cdot \bold x'
 - \bold k''\cdot \bold x'', \tag B.4 $$
$$ A = \text{Tr}\,(C \hat x)^2 = x_k x^k - \tfrac 12 x_{rs} x^{rs} =
 -(x^0)^2 + (\bold x)^2 - (\bold x')^2  + (\bold x'')^2, \tag B.5 $$
$$ -s^2 = \text{Tr}\,(\hat k C^{-1})^2 = k_k k^k - \tfrac 12 k_{rs} k^{rs} =
 -(k^0)^2 + (\bold k)^2 - (\bold k')^2  + (\bold k'')^2, \tag B.6 $$
$$B = 16 \det \hat x= ((x^0)^2 -(\bold x)^2- (\bold x')^2 - (\bold x'')^2)^2 -
\tag B.7 \nopagebreak  $$
$$ -4 \|\bold x \times \bold x'\|^2 -4 \|\bold x' \times \bold x''\|^2
- 4 \|\bold x'' \times \bold x\|^2
- 8 x^0  \bold x'' \cdot \bold x'\times \bold x  $$
and a similar formula for $\det \hat k$.  Note that the parameter $s$
which appears in eqs.\ (B.6) and (9.10) was indicated by $2 s$ in ref.\ 30.

A vector of $\Cal T$ belongs
to $\Cal T^+$ when $x^0$ is larger or equal to the largest root
of the equation $\det \hat x = 0$ and a similar statement holds for
$\Cal T^{*+}$. For $\bold k = 0$ and $k^0 = \mu > 0$, we have
$$ 16 \det \hat k= (\mu^2 - (\bold k')^2 - (\bold k'')^2)^2
 -4 \|\bold k' \times \bold k''\|^2 \tag B.8 $$
and the condition for belonging to $\Cal T^{*+}$ is just eq.\ (9.1).

The quantities $\pm \lambda_{1,2}$ introduced in Sec. IX are the
eigenvalues of the matrix $2 C \hat x$ (the factor $2$ was not present in
 ref.\ 30). They are given by eq.\ (9.27),
where $A$ and $B$ are given by eqs.\ (B.5) and (B.7). If we assume eqs.\
(4.5) and (8.27), we get eq.\ (9.28).  It has been shown in ref.\ 30 that
in a flat-space theory invariant under $Sp(4, \bold R)$ the commutator
vanishes unless one of the quantities $\lambda_{1, 2}^2$ is real negative.
This means that the commutator vanishes if $A^2 - B < 0$ or if
$ A > 0, \, B> 0$. These conditions are satisfied if $\sigma^2 < 0$ or if eq.\
(9.3) holds.

\vfill \eject

\centerline{REFERENCES}
\medskip

\item{$^{1}$} F. Lur\c cat,
Physics {\bf 1}, 95 (1964).

\item{$^{2}$} M. Toller,
Nuovo Cimento {\bf B 64}, 471 (1981).

\item{$^{3}$} M. Toller,
Inter. Journ. Theor. Phys. {\bf 29}, 963 (1990).

\item{$^{4}$} E. R. Caianiello,
Lett. Nuovo Cimento {\bf 32}, 65 (1981).

\item{$^{5}$} H. E. Brandt,
Lett. Nuovo Cimento {\bf 38}, 522 (1983).

\item{$^{6}$} H. E. Brandt,
{\it Proceedings of the Fifth Marcel Grossmann Meeting}
edited by D. G. Blair and M. J. Buckingham, (World Scientific,
Singapore, 1989), pp. 777-785.

\item{$^{7}$} M. Toller,
Nuovo Cimento {\bf B 102}, 261 (1988).

\item{$^{8}$} M. Toller,                                                        

Nuovo Cimento {\bf B 44}, 67 (1978).

\item{$^{9}$} M. Toller and L. Vanzo,
Lett. Nuovo Cimento {\bf 22}, 345 (1979).

\item{$^{10}$} G. Cognola, R. Soldati, M. Toller, L. Vanzo and S. Zerbini,
Nuovo Cimento {\bf B 54}, 325 (1979).

\item{$^{11}$} In some preceeding articles$^{2, 7-10}$ the
fields $A_{\alpha}$ were considered as the generators of the left 
translations.
We have changed this convention in order to be consistent with other
well established conventions.

\item{$^{12}$} S. Kobayashi and K. Nomizu,
{\it Foundations of Differential Geometry}
(Wiley, New York, 1969).                                                       

\item{$^{13}$} Y. Ne'eman and T. Regge,
Riv. Nuovo Cimento {\bf 1, n. 5}, 1 (1978).

\item{$^{14}$} P. K. Smrz,
J. Math. Phys. {\bf 19}, 2085 (1978).

\item{$^{15}$} G. Cognola, R. Soldati, L. Vanzo and S. Zerbini,
J. Math. Phys. {\bf 20}, 2613 (1979).

\item{$^{16}$} V. F. Mukhanov,
JETP Lett. {\bf 44}, 63 (1986).

\item{$^{17}$} V. Bargmann,
Ann. Math. {\bf 59}, 1 (1954).

\item{$^{18}$} For the groups we shall consider, projective representations
are obtained from unitary representations of the universal covering
group$^{17}$.

\item{$^{19}$} R. F. Streater and A. S. Wightman,
{\it PCT, Spin and Statistics, and All That}
(Benjamin, New York, 1964).

\item{$^{20}$} R. Jost,
{\it The General Theory of Quantized Fields}
(American Math. Soc., Providence, Rhode Island, 1965).

\item{$^{21}$} E. P. Wigner,
Ann\. of Math\. {\bf 40}, 149 (1939).

\item{$^{22}$} M. A. Naimark,
{\it Linear Representations of the Lorentz Group}
(Pergamon Press, Oxford, 1964).

\item{$^{23}$} I. M. Gel'fand, M. I. Graev and N. Ya. Vilenkin,
{\it Generalized Functions, Vol. 5}
(Academic Press, New York, 1966).

\item{$^{24}$} A. Sciarrino and M. Toller,
J. Math. Phys. {\bf 8}, 1252 (1967).

\item{$^{25}$} W. R\"uhl,
{\it The Lorentz Group and Harmonic Analysis}
(Benjamin, New York, 1970).

\item{$^{26}$} N. Ya. Vilenkin,
{\it Fonctions sp\'eciales et th\'eorie de la repr\'esentation des groupes}
(Dunod, Paris, 1969).

\item{$^{27}$} G. N. Watson,
{\it Theory of Bessel Functions}
(Cambridge University Press, 1966).

\item{$^{28}$} S. Str\"om,
Arkiv Fysik {\bf 34}, 215, (1967).

\item{$^{29}$} A. Erd\'elyi, W. Magnus, F. Oberhettinger and F. G. Tricomi,
{\it Tables of Integral Transforms}
(McGraw Hill, New York, 1954).

\item{$^{30}$} M. Toller,
Nuovo Cimento {\bf B 108}, 245 (1993).

\item{$^{31}$} M. Toller,
J. Math. Phys. {\bf 24}, 613 (1983).

\bye